\documentclass{article}

\usepackage{graphicx} 
\usepackage{amsmath}
\usepackage{bm}
\usepackage{amssymb}
\usepackage{mathtools}
\usepackage{MnSymbol}
\usepackage{booktabs}
\usepackage{multirow}
\usepackage{threeparttable}
\usepackage{xcolor}
\usepackage{soul}
\usepackage{makecell}
\usepackage{soul}
\usepackage{tabularx} 
\usepackage{caption}  
\usepackage{authblk} 
\usepackage{float}
\usepackage{xurl}
\usepackage{titlesec}

\usepackage[
  backend=biber,
  style=numeric-comp,
  sorting=none,
  giveninits=true,
  maxnames=99
]{biblatex}

\DeclareNameAlias{author}{family-given}
\DeclareNameAlias{editor}{family-given}

\AtEveryBibitem{\clearfield{number}}

\renewbibmacro{in:}{}

\usepackage{comment}

\usepackage[a4paper, left=15mm, right=15mm, top=20mm, bottom=20mm]{geometry}

\title{Harnessing Machine Learning for Hybrid Constitutive Modelling of Viscoelastic Fluid Flows in Computational Rheology}

\date{}

\author[1]{J.L. Cummings}
\author[2,3]{C. Fernandes}
\author[4]{F. Dong}
\author[3,5]{M.A. Alves}
\author[1]{M.S.N. Oliveira}

\affil[1]{James Weir Fluids Laboratory, Department of Mechanical and Aerospace Engineering, University of Strathclyde, Glasgow, UK}
\affil[2]{Department of Mechanical Engineering, Faculty of Engineering, University of Porto, Portugal}
\affil[3]{CEFT, ALiCE, Faculty of Engineering, University of Porto, Portugal}
\affil[4]{Department of Computer and Information Sciences, University of Strathclyde, Glasgow, UK}
\affil[5]{Department of Chemical and Biological Engineering, Faculty of Engineering, University of Porto, Portugal}

\begin{document}

\maketitle

\section{Abstract}

Recent advances in data-driven modelling have highlighted the potential of hybrid approaches which combine Tensor Basis Neural Networks (TBNN) with Universal Differential Equations (UDE) to discover frame-invariant, non-linear viscoelastic constitutive models. These hybrid models enable the creation of digital twins for complex viscoelastic fluids, offering direct transferability to computational fluid dynamics simulations. In this work, we introduce a reduced-dimensional tensor basis formulation that enhances both the physical consistency of the learned representations with respect to the training data and the numerical stability of subsequent simulations. The UDE architecture is embedded into an open-source finite volume solver in which the constitutive response is generated dynamically at runtime based on local fluid flow conditions
. Training on synthetic datasets generated using a range of  well established viscoelastic models in oscillatory shear flows alone, the performance of the resulting UDEs is evaluated under extrapolation to unseen conditions and flow-types. These include deploying the UDEs in viscometric extensional flows as well as 
2D and 3D benchmark flows, such as the 4:1 sudden contraction and cross-slot, providing a quantitative analysis of their capabilities, limitations and failure modes. 
The proposed reduced-basis framework enables data-efficient discovery of frame-invariant constitutive models that generalise beyond their training regime, capturing key flow features such as the onset and growth of flow-induced elastic instabilities in strong extensional flows even though trained solely on shear data. Quantitative accuracy decreases as extrapolation increases, but incorporating first normal stress difference information further improves quantitative accuracy and extends predictive fidelity to higher Deborah numbers.

\section{Introduction}

Accurate numerical predictions of viscoelastic flows hinge on robust constitutive laws that quantitatively relate the local deformation history to the evolution of the extra-stress tensor. Since the seminal work of Oldroyd \cite{oldroyd1950formulation}, which introduced the concept of frame-invariant upper and lower convected derivatives, a wide range of constitutive models has been proposed to describe the evolution of the stress tensor of complex materials under arbitrary deformation protocols and histories. Classical models such as the Oldroyd-B \cite{oldroyd1950formulation}, Giesekus \cite{giesekus1982simple}, Phan-Thien Tanner (PTT) \cite{thien1977new}, as well as fractional viscoelastic models \cite{schiessel1995generalized}, have been developed to encode key viscoelastic features, such as non-linear elasticity, relaxation spectra, and memory effects. 

The advent of computational methods for solving viscoelastic fluid flow problems \cite{oliveira1998numerical}, and in recent years the increasing popularity and trust in open-source codes such as rheoTool \cite{pimenta_stabilization_2017} or Basilisk \cite{Popinet} have offered new opportunities to better understand the behaviour of complex fluid flows. However, a rheologist wishing to perform CFD-based flow simulations must first select a suitable constitutive model and then calibrate its parameters, typically based on limited material property data. This procedure is often guided by the intuition and prior experience of the analyst, and the choice of model has far reaching implications in terms of the resulting predictions. Although classical models provide valuable insight into the complex flow behaviour of such fluids, their prescribed functional forms invariably impose simplifications which limit their ability to reproduce the full, rich dynamics observed in real flows. A well-known example is the difficulty in predicting (even qualitatively) the correct pressure drop of Boger fluids, such as dilute polymeric solutions, flowing through contraction/expansion geometries even when the underlying kinematics are accurately captured \cite{alves_benchmark_2003}. 

\begin{table}[htbp]
\begin{threeparttable}
\centering
\caption{Overview of Machine Learning Approaches for Constitutive Modelling and Computational Rheology}
\renewcommand{\arraystretch}{1.3} 
\setlength{\tabcolsep}{6pt}       
\small
\begin{tabular}{p{3cm}p{2.2cm}p{2.5cm}p{6cm}}
\hline
\textbf{Method / Paper} 
    & \textbf{Approach} 
    & \textbf{Physics Informed (PI)/Data Driven} 
    & \textbf{Notes} \\
\hline
nn-PINNs \cite{mahmoudabadbozchelou_nn-pinns_2022}       & DNN & PI & Extends classical PINNs to include viscoelastic fluids, considering linear physics-informed terms. \\
RheologyNet \cite{zhang2023rheologynet}                 & DNN & PI & Developed specifically for cementitious materials. Very good accuracy compared to other computational approaches.  \\
ViscoelasticNet \cite{thakur_viscoelasticnet_nodate}   & DNN & PI + Data & Automates model selection and parameterisation, also able to generate stress fields purely from velocity data. \\
RhINNs \cite{saadat_datadriven_nodate}                  & DNN & PI + Data & Automated model selection and parameterisation approach. \\
ConNN \cite{jin_data-driven_nodate}                     & RNN & Data & The recurrent unit functions as a direct surrogate for fluid memory. \\
Chen \cite{chen_recurrent_nodate}                       & RNN & Data & Analyses the capabilities of the "long short term memory" unit in addition to the recurrent unit. \\
NARX NN \cite{mishra_one_2025}                          & DNN + Feedback Loop & Data & Including history sensitivity gives significantly improved performance over classical structural kinetic models.  \\
RheOFormer \cite{saberi2025rheoformer}                 & Transformer & Data & Full CFD surrogate able to accurately predict time-dependent multi-dimensional flows. \\
Rheo-SINDy \cite{sato_rheo-sindy_nodate}               & Symbolic Regression & Data & Can retain frame invariance if candidate function library is judiciously selected.  \\
Shanbhag \& Erlebacher \cite{shanbhag2024sparse}       & Symbolic Regression & Data & Uses tensor basis functions to guarantee frame-invariance. \\
GENERIC-Guided \cite{simavilla2024hammering}           & DNN & PI + Data & Directly applicable to CFD methods,  obtains extra-stress tensor via automatic differentiation approach and satisfies entropy considerations by design. \\
Darves-Blanc et al. \cite{darves4748606cfd}            & DNN & Data & CFD implementation of ML-based viscosity function. Frame-invariance not guaranteed; instead achieved by augmenting training data with rotations and reflections. \\
RUDE \cite{lennon2023scientific}                       & TBNN & Data & Frame-invariant and non-linear correction to a base model. \\
Differentiable CFD \cite{sunol2025learning}            & CFD + TBNN & Data & Learns a TBNN representation with end-to-end differentiable CFD; projects learned model onto existing classical models via LAOS-based approach.  \\
Sparse Regression + RhINNs \cite{mahmoudabadbozchelou_unbiased_2024} & NN + Symbolic Regression & PI + Data & Can retain frame invariance if candidate function library is judiciously selected.   \\
FNO \cite{mangal2025learning}                           & FNO & & Exploits the function-to-function mapping of FNOs to learn constitutive model families. \\
MFNN \cite{mahmoudabadbozchelou_data-driven_nodate}    & MFNN & PI + Data & Can learn to relate physicochemical parameters to model behaviour using a combination of low-fidelity synthetic data and high-fidelity experimental data. \\
John et al. \cite{john_machine_2024}                   & Random Forest & Data & Automated model selection and parameterisation via Chevyshev spectra analysis. \\
RhiGNet \cite{mahmoudabadbozchelou2022digital}        & Graph NN & PI + Data& Blends graph NNs with a multi-fidelity approach to create digital twins for rheometric protocols.  \\
\hline
\end{tabular}

\begin{tablenotes}
\footnotesize
\item DNN: Deep Neural Network; RNN: Recurrent Neural Network; FNO: Fourier Neural Operator; MFNN: Multi-Fidelity Neural Network
\end{tablenotes}

\end{threeparttable}
\end{table}

These challenges have motivated the use of machine learning (ML) techniques, whose capabilities extend far beyond the automation and refinement of model parameterisation \cite{saadat_datadriven_nodate, thakur_viscoelasticnet_nodate, john_machine_2024}. ML enables the development of alternative model closures, formulation of new constitutive models (e.g., RheoSINDy \cite{sato_rheo-sindy_nodate}) and even unlocks new model paradigms for the prediction of complex rheological behaviour. 

More generally, ML approaches have been used in non-Newtonian fluid dynamics to bypass conventional CFD model approaches and directly predict the stress response of complex fluids. This has been done by considering classical physics-informed neural networks (PINNs), in which the pertinent constitutive model enters the physics-informed loss component \cite{zhang2023rheologynet, mahmoudabadbozchelou_nn-pinns_2022}, or by directly incorporating complex memory effects and history-dependent behaviour into the ML model topology \cite{jin_data-driven_nodate, chen_recurrent_nodate}.  Recent approaches have adapted the so-called “attention” mechanism of transformers \cite{zhao2023pinnsformer, saberi2025rheoformer}, originally designed for natural language generation \cite{vaswani2017attention}, to move away from the point-wise predictions of classical PINNs and proactively account for the contextual importance of history-dependent temporal structures.

Other approaches in ML have been adapted for rheological modelling, with Table 1 presenting a non-exhaustive summary of key contributions; Fourier neural operators (FNOs) \cite{li2020fourier, mangal2025learning}, unlike classical PINNs, learn mappings between function spaces, allowing them to represent not only a single partial differential equation (PDE) but a family of PDEs with varying parameters or boundary conditions. Multi-fidelity neural networks (MFNNs) \cite{meng2020composite, mahmoudabadbozchelou_data-driven_nodate} exploit the negligible cost of generating synthetic “low-fidelity” data compared to high-fidelity and high cost experimental data. In the context of rheology, by first relating material characteristics (such as colloid fraction and sample age) to a simple classical rheological model using a low-fidelity NN, basic physical intuition is established which is then fine-tuned using a high-fidelity NN trained on experimental data. 

Although many ML architectures have been proposed for predicting stress responses, far fewer works blend ML constitutive modelling and CFD simulations of viscoelastic fluids. Notable exceptions employ hybrid frameworks (e.g.,  \cite{lennon2023scientific, sato2025multiscale, sunol2025learning, simavilla2024hammering, darves4748606cfd}) that integrate ML-based constitutive modelling into CFD simulations of viscoelastic fluids. A key challenge for any model intended for CFD deployment is to preserve frame invariance, ensuring that the model remains valid under large deformations. In contrast, hybrid ML approaches for \textit{turbulence} modelling in CFD have attracted considerable attention \cite{majchrzak2023survey}. These methods are conceptually similar to viscoelastic fluid modelling because both require solving additional evolution equations and constructing appropriate model closures.

One such hybrid framework is the Rheological Universal Differential Equations (RUDE) approach proposed by Lennon et al. \cite{lennon2023scientific}, which we further develop and extend in the present work. RUDE augments a classical constitutive law - specifically the upper-convected Maxwell model in the original formulation of \cite{lennon2023scientific} - with a neural-network correction term that learns the missing physics responsible for non-linear viscoelastic behaviour. This data-driven term is trained directly on material-property data, while the underlying classical structure ensures that frame invariance is preserved. Ideally, such hybrid frameworks should be trainable from limited datasets and should be \textit{transferable}, enabling accurate predictions using CFD simulations across a wide range of previously unseen flow types and conditions.

More specifically, the RUDE framework adapts the tensor basis neural network (TBNN) approach, originally proposed by Ling et al. \cite{ling_reynolds_2016} in the context of closing the Reynolds stress tensor in turbulence modelling. This approach can be described as physics-\textit{guided}, as physical principles are embedded in the learning problem to improve the inductive bias, but ultimately the training procedure is purely data-driven and does not directly enforce known physics in the loss function optimisation as in classical physics-\textit{informed} approaches.  In the TBNN approach, the neural network outputs are projected onto a set of so-called ``integrity basis" tensors, which are constructed according to the canonical assumption that stress and strain-rate are the primary macroscopic flow parameters influencing the evolution of stress. 
The work on RUDEs was extended by Rodrigues et al. \cite{rodrigues2025finding}, going beyond the Giesekus model and considering synthetic data generated by other non-linear viscoelastic models such as the Johnson-Segalman and PTT models. 
Similarly, Shanbhag and Erlebacher \cite{shanbhag2024sparse} adapt the tensor-basis approach from a TBNN to sparse symbolic regression \cite{brunton2016discovering} which has the key advantage of arriving at an easily interpretable closed-form constitutive model. 

By projecting its outputs onto a frame-invariant tensor basis, the RUDE framework inhibits the expressiveness of the neural network in return for portability to new flow-types, geometries and simulation environments. 
The architecture has been successfully deployed in planar start-up shear flow \cite{lennon2023scientific, rodrigues2025finding}, not seen during training, and the proof of concept CFD deployment of a RUDE trained on a Giesekus fluid in a planar contraction at a specific Weissenberg number ($Wi = 1$) \cite{lennon2023scientific} 
is promising. This portability to a scenario involving extensional flow is particularly encouraging since extensional flows can elicit a different microstructural response within a complex fluid that is not necessarily readily inferred purely from observations in shear flow. However, as we shall observe in the forthcoming analysis, the Giesekus model is a particular case in which the TBNN can effectively recover the true constitutive model. As will be demonstrated, this level of model recovery is an exception rather than the rule. The question therefore is how well a trained UDE can perform under deformation histories comprising extensional flows and mixed kinematics when the training is restricted to shear flows alone. Here, we address this question, which is crucial for assessing the practical capabilities of the framework. Further, we examine whether and to what extent providing additional information about the evolution of the stress tensor (such as the first normal stress difference) during training improves the regression quality. 

Another key issue of unique importance to hybrid ML constitutive models in CFD simulations we address is related to numerical stability. CFD simulations of viscoelastic fluids are remarkably demanding, suffering from the well known ``high Weissenberg number problem" \cite{keunings1986high} unless specific precautions such as adopting the log-conformation approach are taken \cite{afonso_log-conformation_2009}. Even so, numerical stability is significantly influenced by the characteristics of the viscoelastic constitutive model employed. For example, due to its unbounded extensional viscosity, it is much more challenging to obtain convergence in simulations of UCM and Oldroyd-B fluids than with simplified Phan-Thien-Tanner (sPTT) fluids \cite{alves2021numerical}. The real-world practicality of a deployed UDE is contingent on its ability to reliably converge in highly elastic flow regimes, and we address this gap by identifying an apparent instability mechanism in the original RUDE framework and introduce a revised formulation which significantly improves stability in high $Wi$ flow regimes.

The remainder of the paper is structured as follows: in Section \ref{sec:Methodology}, the methodology is introduced, followed by the training pipeline and the generation of the datasets used during training in Section \ref{sec:Training}, and the deployment of the trained model in the context of CFD simulations in Section \ref{sec:DeployCFD}. Section \ref{sec:results} provides a comprehensive analysis of the behaviour of the trained UDEs, including direct observations of the learned model behaviours associated with the input space encountered during training and the analysis of material properties such as the steady shear and extensional viscosity, and the first normal stress difference coefficients. We then consider the behaviour of trained UDEs in deployment to CFD simulations, discussing improvements of the current formulation in terms of the numerical instabilities relative to previously proposed three-dimensional formulations, and assessing performance in a range of flow geometries and conditions.

\section{Universal Differential Equation Framework and Tensor Basis Neural Networks} \label{sec:Methodology}

In the present work, we employ the UDE framework proposed by Rackauckas et al. \cite{rackauckas_universal_nodate}, which in turn builds on the fundamental work of Chen et al. \cite{chen2018neural} introducing Neural Ordinary Differential Equations (neural ODEs). The  key idea underpinning neural ODEs is that they aim to learn a latent derivative function such that, when integrated from given initial states, the resulting solution matches the observed states. Typically, this is achieved by incorporating the neural network into a numerical ODE solver; the neural network predicts the derivative at any point and the solver integrates these predictions to approximate the trajectory of the function. As universal function approximators \cite{hornik1989multilayer}, neural networks can effectively serve as surrogate models for the underlying derivative function. Unlike a classical supervised training structure, the NN outputs the time derivative and the integrated state is then compared to the ground truth. Since the outputs of the NN parameterise an ODE, an integration step must be performed; thus in the standard neural ODE formulation the network does not directly output the observed states but instead a derivative field that reproduces them when integrated. A fundamental example of a neural ODE is shown in Eq. \ref{eq:1}, where the temporal derivative of an output state $y$ at a spatiotemporal point $(\bm{x}, t)$ is given by a derivative function $f_\theta$ described by a deep neural network whose inputs are $(\bm{x}, y(\bm{x},t), t)$ and with weights and biases represented by $\theta$: 

\begin{equation} \label{eq:1}
\frac{\partial y(\bm{x},t)}{\partial t} = f_{\theta}(\bm{x}, y(\bm{x}, t), t; \theta).
\end{equation}

The essential augmentation introduced in the UDE is the incorporation of prior knowledge such as empirical or mechanistic models. These are included as analytical terms in the ODE, and are believed to (at least in part) describe the behaviour of a system. With this addition, the role of the neural network is no longer to fully parameterise the hidden derivative of the system, but rather to tune the general behaviour introduced by the prior knowledge terms. This is shown in Eq. \ref{eq:2}, including $s$ analytical terms: 

\begin{equation}\label{eq:2}
\frac{\partial y(\bm{x},t)}{\partial t} = \sum_{i=1}^{s}{f_{m, i}(\bm{x},y(\bm{x}, t), t)} +f_{\theta}(\bm{x},y(\bm{x}, t), t; \theta)
\end{equation}

\noindent where $f_{m,i}$ refers to the $i$th modelling term representing a-priori knowledge.
Following the framework of Lennon et al. \cite{lennon2023scientific}, we adopt a UDE framework and, by default, employ the upper-convected Maxwell (UCM) model to represent prior knowledge. The UCM model is a simple differential-type and frame-invariant model able to qualitatively capture viscoelastic behaviour, and many complex viscoelastic models simplify to the UCM model in the limit of appropriate model parameters (e.g. for the Giesekus model when $\alpha \rightarrow 0$) or in the linear, small rate of deformation limit. The constitutive equation can then be expressed as:

\begin{equation} \underbrace{\boldsymbol{\sigma} + \lambda \stackrel{\kern0.0005em\smalltriangledown}{\boldsymbol{\sigma}}}_{\text{Base model (UCM)}} +  \underbrace{\frac{\eta_p}{\lambda} \bm{\widetilde{F}}(\boldsymbol{\sigma}, \dot{\bm{\gamma}}; \theta)}_{\text{Tensor Basis Neural Network}} = \underbrace{\eta_p \dot{\bm{\gamma}}}_{\textnormal{Base Model (UCM)}} 
\end{equation}
where $\boldsymbol{\sigma}$ is the viscoelastic extra-stress tensor, $\lambda$ is the relaxation time of the fluid, $\eta_p$ is the polymeric viscosity and $\dot{\bm{\gamma}}$ is the strain-rate tensor given by $\dot{\bm{\gamma}} = \nabla \boldsymbol{u} + (\nabla \boldsymbol{u})^\textnormal{T}$. The upper-convected derivative of the viscoelastic extra-stress tensor is given by 
$\stackrel{\kern0.0005em\smalltriangledown}{\boldsymbol{\sigma}} = \frac{\partial \boldsymbol{\sigma}}{\partial t} + \boldsymbol{u}\cdot \nabla \boldsymbol{\sigma} - \left( \nabla \boldsymbol{u}^\textnormal{T}\cdot\boldsymbol{\sigma} + \boldsymbol{\sigma}\cdot\nabla \boldsymbol{u} \right)$.
The use of a frame-invariant derivative is crucial for accurately modelling behaviour in large deformation flows. 

The non-dimensional $\boldsymbol{\widetilde{F}}$ term in Eq. 3 represents the neural network contribution to the differential equation, and in this case a tensor basis neural network (TBNN) is employed. The motivation underpinning the tensor basis neural network is the need to retain Galilean invariance, and in the present work, full frame objectivity under rotations also. An invariant formulation can be ensured by expressing 
$\bm{\widetilde{F}}$ as a linear combination of \textit{basis tensors}, each multiplied by scalar functions of the corresponding \textit{invariants}. The basis tensors are objective tensorial combinations constructed to satisfy frame-invariance, while the associated integrity basis consists of the minimal, complete set of scalar invariants and tensor basis functions required to represent any admissible constitutive relation. In this framework, all possible tensorial relationships are captured as linear combinations of the basis tensors weighted by functions of the invariants; in the context of a macroscopic rheological model where the stress and strain-rate are the only tensorial flow quantities influencing the stress evolution, the integrity basis is a complete description of all of the possible interaction of the stress and strain-rate. $\bm{\widetilde{F}}$ is weighted via a linear combination of the basis tensors:

\begin{equation}
\bm{\widetilde{F}} =\sum_{n=1}^{N} g_n(\bm{\sigma},\dot{\bm{\gamma}} ;\theta) \bm{T}_n
\end{equation} 

\noindent where $g_n(\bm{\sigma},\dot{\bm{\gamma}}; \theta)$ is the scalar output of the neural network at neuron $n$ of the output layer which is associated with basis tensor $\bm{T}_n$.

In previous works, the set of integrity basis tensors has been derived following the work of Spencer and Rivlin \cite{spencer1958theory} assuming generality in three dimensions (i.e. $\boldsymbol{\sigma}$ and $\dot{\bm{\gamma}}$ are second order tensors over $\mathbb{R}^3$), yielding a set of nine tensors. A re-derivation by Rodrigues et al. \cite{rodrigues2025finding} based on the work of Smith \cite{smith1971isotropic}, yields a reduced set of eight tensors, also assuming generality in three dimensions but omitting the quartic term (i.e. the term in $ \bm{\sigma} \cdot \bm{\sigma}  \cdot \dot{\bm{\gamma}} \cdot \dot{\bm{\gamma}} + \dot{\bm{\gamma}} \cdot \dot{\bm{\gamma}} \cdot \bm{\sigma} \cdot \bm{\sigma}  $). There has been some debate in the literature regarding the necessity of this quartic term but Kamrin and Govindjee \cite{kamrin2025clarifying} recently showed that the quartic term is not required. The reduced set of eight basis tensors is sufficient to represent all isotropic tensor-valued functions of symmetric second-order tensors in 3D.

In the present work, we adopt a two-dimensional formulation (i.e. assuming that $\boldsymbol{\sigma}$ and $\dot{\bm{\gamma}}$ are second order tensors over $\mathbb{R}^2$) as the foundation for our integrity basis. This choice is motivated by two considerations. Firstly, the rheometric flow protocols adopted in training are two-dimensional, and so the set of invariants and basis tensors used in the three-dimensional formulation is no longer fully informative.  In planar shear flows, certain invariants which were proposed as inputs to the neural network (and are required to fully capture 3D interactions between stress and strain-rate) vanish (e.g. $\textnormal{tr}(\dot{\bm{\gamma}} \cdot \dot{\bm{\gamma}} \cdot \dot{\bm{\gamma}}) $), and clearly a model purely trained on such cases cannot be expected to generalise to flows where these invariants are non-zero. This was partially recognised by Shanbhag and Erlebacher \cite{shanbhag2024sparse}, who pointed out that several of the invariants employed in their sparse regression algorithm are either zero or linear combinations of other invariants. Less obviously, when training purely on 2D flows (whether pure shear or otherwise), all invariants based on three-dimensional interactions, though they can appear to be non-zero during training depending on the flow kinematics governing the training data, are not truly independent because they
reduce to linear combinations of two-dimensional terms and, therefore, do not provide any new information about the flow. In practice, this means that although these features may nominally be activated during training, they cannot encode distinct three-dimensional flow physics. When such a model is then deployed in a genuinely three-dimensional flow, it is likely to misrepresent the underlying rheology since it has never encountered or learned to distinguish three-dimensional interactions in the input space. Addressing this limitation would require a fundamental redesign of the training procedure to accommodate genuinely three-dimensional flows. The training pipeline presented here relies on the stress field being homogeneous in the training data; although this can easily be realised in any rotational rheometer, producing an analogously homogeneous stress field in three dimensions is extremely difficult to realise experimentally across a finite sample, even though it can be well-defined theoretically. We therefore argue that three-dimensional interactions should be avoided when the training data are derived exclusively from 2D flows. Secondly, the two-dimensional integrity basis offers significantly improved numerical stability during deployment to CFD simulations. As will be discussed in greater detail in Section \ref{subsec:cfdeployment}, our tests in deploying a trained network to a CFD simulation have demonstrated numerical instability directly associated with the higher order terms 
present in the three-dimensional formulation
. It is critical to emphasise that the two-dimensionally motivated integrity basis is not incompatible with fully three-dimensional CFD, nor are third-order constitutive model terms necessarily a requirement for simulating three-dimensional viscoelastic fluid flows. Indeed, it is worthwhile to note that this choice restricts the model class to a subset of all 3D frame invariant constitutive relations, but still produces an objective model which is consistent with the functional form of most commonly used constitutive models, which rarely employ the higher-order tensor products present in the full basis. The validity of the two-dimensional integrity basis in a 3D CFD simulation is explicitly demonstrated in Section \ref{subsubsec:3D}.

According to the Cayley-Hamilton theorem, in two dimensions, a minimal tensor basis is the following:

\begin{equation}
\mathcal{T} = \{ \bm{I}, \bm{\sigma}, \dot{\bm{\gamma}} \}
\end{equation}

In the present work, we choose to move away from minimal mathematical completeness towards a set of basis tensors which enhances physical expressiveness based on conventional rheological modelling principles. This is done by including key non-linear interactions in the basis tensor representation: here, we opt to include $\bm{\sigma} \cdot \dot{\bm{\gamma}} + \dot{\bm{\gamma}} \cdot \bm{\sigma}$ in the 
set of basis tensors because it directly encodes coupled stress-strain-rate interactions. This term introduces a corotational type derivative which is essential for describing slippage as in the Johnson-Segalman or the Phan-Thien-Tanner models with $\xi \neq 0$.   
Similarly, we opt to replace the identity tensor with $\bm{\sigma} \cdot \bm{\sigma}$ on account of two factors. Firstly, the inclusion of the identity tensor introduces the potential for problematic behaviour if it is unconstrained. For instance, a poor regression could produce a non-zero coefficient for the identity tensor when an inference is generated for a fluid in a rest state (i.e. $\bm{\sigma} = \dot{\bm{\gamma}} = \bm{0}$). In this case, the inferred constitutive model would predict a non-zero rate of change of stress even for a completely quiescent fluid, which is clearly unphysical. We anticipate that a constraint mechanism, analogous to the approach to enforcement of boundary conditions commonly employed in PINNs, may be required in the form of a ``quiescence penalty'' in order to properly navigate this issue. Soft enforcement, for instance, may be implemented by asking the TBNN at each training epoch to generate a constitutive model for a fluid at rest state (where all of the invariants are zero) and adding a term in the loss function which penalises the coefficient in $\bm{I}$ which is output by the TBNN. However, this approach introduces an additional tuning step because an appropriate scaling coefficient must be empirically determined \cite{berrone2023enforcing}. Secondly, the term  $\bm{\sigma} \cdot \bm{\sigma}$, commonly recognised as a Giesekus-like contribution representing flow anisotropy, is more common than the isotropic behaviour modelled by the identity tensor, which can appear in constitutive equations derived from Doi-Edwards tube type models \cite{Doi_Edwards_1986}. Although, in principle, a TBNN constructed with a minimal integrity basis can represent the interactions of $\bm{\sigma} \cdot \bm{\sigma} $, in practice this requires the scalar coefficient functions to learn highly non-linear combinations of invariants in order to recover meaningful underlying couplings. This places a significant burden on the learning process. By explicitly including $\bm{\sigma} \cdot \bm{\sigma} $ and $\bm{\sigma} \cdot \dot{\bm{\gamma}} + \dot{\bm{\gamma}} \cdot \bm{\sigma}$ in the basis, these non-linear interactions are directly encoded at the tensorial level. This reduces the complexity required of the learned scalar functions and provides a more favourable inductive bias, and also ensures that the representation explicitly reflects established non-linear rheological mechanisms. According to the Cayley-Hamilton theorem $\bm{\sigma} \cdot \bm{\sigma} = f(\bm{I}, \bm{\sigma})$, and so the exchange of tensors does not prevent the emergence of an isotropic contribution (i.e. a term in $\bm{I}$) so long as $\bm{\sigma}$ remains in the integrity basis set, except in the limit when $\textnormal{det}(\bm{\sigma}) = 0$ (or $\bm{\sigma} = \bm{0}$) which is not problematic given that the TBNN goal is to learn corrections from the non-zero stress/strain-rate state. 

Therefore, the final set of integrity basis tensors employed in this work is the following:

\begin{equation}
\mathcal{T} = \{ \bm{\sigma}, \dot{\bm{\gamma}}, \bm{\sigma} \cdot \bm{\sigma} ,\left( \bm{\sigma} \cdot \dot{\bm{\gamma}} + \dot{\bm{\gamma}} \cdot \bm{\sigma} \right) \}
\end{equation} \label{eq:6}

\noindent which has been selected for physical interpretability as well as compatibility with classical constitutive equations, rather than the minimal complete basis in 2D, whose behaviour is more challenging to constrain and forces the TBNN to identify more complex representations if it is to accurately capture well-known non-linear stress-strain-rate coupling effects.

In the present work, TBNN models employing eight and nine basis tensors, as used in \cite{rodrigues2025finding} and \cite{lennon2023scientific}, respectively, are also implemented in the framework of the same code for comparison:

\begin{equation}
\mathcal{T}_8 = \{ \bm{I},  \bm{\sigma}, \dot{\bm{\gamma}}, \bm{\sigma} \cdot \bm{\sigma} ,\dot{\bm{\gamma}} \cdot \dot{\bm{\gamma}}, \left( \bm{\sigma} \cdot \dot{\bm{\gamma}} + \dot{\bm{\gamma}} \cdot \bm{\sigma} \right), (\bm{\sigma} \cdot \bm{\sigma} \cdot \dot{\bm{\gamma}} +\dot{\bm{\gamma}} \cdot \bm{\sigma} \cdot \bm{\sigma} ), (\bm{\sigma} \cdot \dot{\bm{\gamma}} \cdot \dot{\bm{\gamma}} +\dot{\bm{\gamma}} \cdot \dot{\bm{\gamma}} \cdot \bm{\sigma} )  \}
\end{equation}\label{eq:7}

\begin{equation}
\mathcal{T}_9 = \{ \bm{I},  \bm{\sigma}, \dot{\bm{\gamma}}, \bm{\sigma} \cdot \bm{\sigma} ,\dot{\bm{\gamma}} \cdot \dot{\bm{\gamma}}, \left( \bm{\sigma} \cdot \dot{\bm{\gamma}} + \dot{\bm{\gamma}} \cdot \bm{\sigma} \right), (\bm{\sigma} \cdot \bm{\sigma} \cdot \dot{\bm{\gamma}} +\dot{\bm{\gamma}} \cdot \bm{\sigma} \cdot \bm{\sigma} ), (\bm{\sigma} \cdot \dot{\bm{\gamma}} \cdot \dot{\bm{\gamma}} +\dot{\bm{\gamma}} \cdot \dot{\bm{\gamma}} \cdot \bm{\sigma} ), (\bm{\sigma} \cdot \bm{\sigma} \cdot\dot{\bm{\gamma}} \cdot \dot{\bm{\gamma}} +\dot{\bm{\gamma}} \cdot \dot{\bm{\gamma}} \cdot \bm{\sigma}  \cdot \bm{\sigma} )  \}
\end{equation}\label{eq:8}

\section{Training Pipeline}
\label{sec:Training}
\subsection{Generation of Training Data}

In the present study, we use synthetic data generated using five well-established viscoelastic constitutive models to train the neural network.
When the base model is the UCM model, the analytical form of the target function $\bm{F} = \frac{\eta_p}{\lambda}\widetilde{\bm{F}}\left( \bm{\sigma}, \dot{\bm{\gamma}} \right)$ is as shown in Table 2. Unless otherwise specified, a solvent viscosity ratio of zero is employed, i.e., the fluid is assumed to have no Newtonian solvent component contributing to the stress state. Although this is consistent with previous works \cite{lennon2023scientific, rodrigues2025finding}, it is not a physically reasonable assumption - we also performed training runs using datasets generated by a fluid with non-zero solvent viscosity and found that it had a negligible influence both on the stability of the training process and on the behaviour learned by the TBNN. This is a natural consequence of the fact that the TBNN only learns to predict the non-linear contribution associated with the evolution in the polymeric stress component.

\begin{table}[H] 
\centering
\renewcommand{\arraystretch}{1.4} 
\begin{threeparttable}
\caption{Target function associated with each of the constitutive models employed in the generation of synthetic data (using UCM as the base model). }
\label{tab:gt-models}
\begin{tabular}{ll}
\toprule
\textbf{Model} & \textbf{$\boldsymbol{F} = \frac{\eta_p}{\lambda}\widetilde{\bm{F}} \left( \boldsymbol{\sigma}, \dot{\bm{\gamma}} \right)$} \\
\midrule
Upper Convected Maxwell/Oldroyd-B & $\bm{0}$ \\
Giesekus & $\frac{\alpha \lambda}{\eta_p} \left( \bm{\sigma} \cdot \bm{\sigma} \right)$ \\
Johnson-Segalman & $\frac{\lambda \xi}{2} \left( \bm{\sigma} \cdot \dot{\bm{\gamma}} 
+ \dot{\bm{\gamma}} \cdot \bm{\sigma} \right)$ \\
Linear/Simplified Phan-Thien-Tanner (L/sPTT) & $ \frac{\lambda \epsilon}{\eta_p} \, \text{tr}(\bm{\sigma}) \, \bm{\sigma} 
+ \frac{\lambda \xi}{2}\left( \bm{\sigma} \cdot \dot{\bm{\gamma}} 
+ \dot{\bm{\gamma}} \cdot \bm{\sigma} \right)$  
($\xi = 0 \rightarrow \text{sPTT}$) \\
Exponential Phan-Thien-Tanner (ePTT) & $\left( \exp\!\left(\tfrac{ \lambda \epsilon}{\eta_p}\text{tr}(\bm{\sigma}) \right) - 1 \right)\bm{\sigma} + \frac{\lambda \xi}{2}\left( \bm{\sigma} \cdot \dot{\bm{\gamma}} 
+ \dot{\bm{\gamma}} \cdot \bm{\sigma} \right) $ \\
\bottomrule
\end{tabular}
\end{threeparttable}
\end{table}
 
Consistent with previous works \cite{lennon2023scientific, rodrigues2025finding}, we choose to employ a combination of small and large amplitude oscillatory shear flows (S/LAOS) as the main deformation protocol to generate the training data for each of the 
ground-truth fluids: 

\begin{equation}
\dot{{\gamma}}(t) = \gamma_0 \omega \textnormal{cos}(\omega t)
\end{equation}\label{eq:9}

\noindent where $\dot{\gamma}$ is the instantaneous imposed shear rate, $\gamma_0$ is the strain amplitude and $\omega$ is the angular frequency.

The choice of LAOS is convenient not only because it is an effective flow protocol for exposing the non-linear rheology of a fluid, but also because S/LAOS flow is homogeneous and is fully described by a single (spatially uniform) stress and strain-rate tensor at each time. Oscillatory shear flows are also frequently used in experimental rheological characterisation, which is crucial for application to real-world fluids.

Sets of between 8 and 12 stress signals (with different combinations of frequency and amplitude) are generated using each ground-truth constitutive model, encompassing both the small and large amplitude regime.
  The datasets span a broad range of flow conditions, with $ 0.1\leq Wi \leq 40$ and $ 0.25 \leq De \leq 2$, with the Deborah ($De$) and Weissenberg ($Wi$) numbers given by $De = \lambda \omega$  and $Wi = \gamma_0De$. We deliberately choose to explore the very large amplitude oscillatory shear regime, even though it is challenging to realise experimentally, because doing so exposes the TBNN to a broader range of inputs during training, which might encourage improved generalisation capabilities when the model faces complex flows where local elastic effects become large.

\subsection{Training Methodology}
In the present work, we use Python 3.12.7 for the training pipeline due to the availability of libraries for neural network training (e.g. PyTorch \cite{paszke2019pytorch} or JAX \cite{frostig2019compiling}), and also because it offers the possibility to embed the neural network into a CFD solver written in C++ (e.g. OpenFOAM) by initialising a Python interpreter within the C++ code, for instance with Pybind \cite{jakob2024pybind11, rodriguez2022general, simavilla2024hammering}. This approach would make other neural network topologies (e.g. convolutional or recurrent networks and transformers) straightforward to embed into the solver in future work because the inference step is outsourced.

The goal of the training process is to optimise the weights and biases of the TBNN in order to produce a constitutive model for the viscoelastic extra-stress tensor such that the resulting predictions match the stress evolution of the target fluid (i.e. ground-truth).

\begin{figure}[htp]
    \centering
    \includegraphics[width=0.75\textwidth]{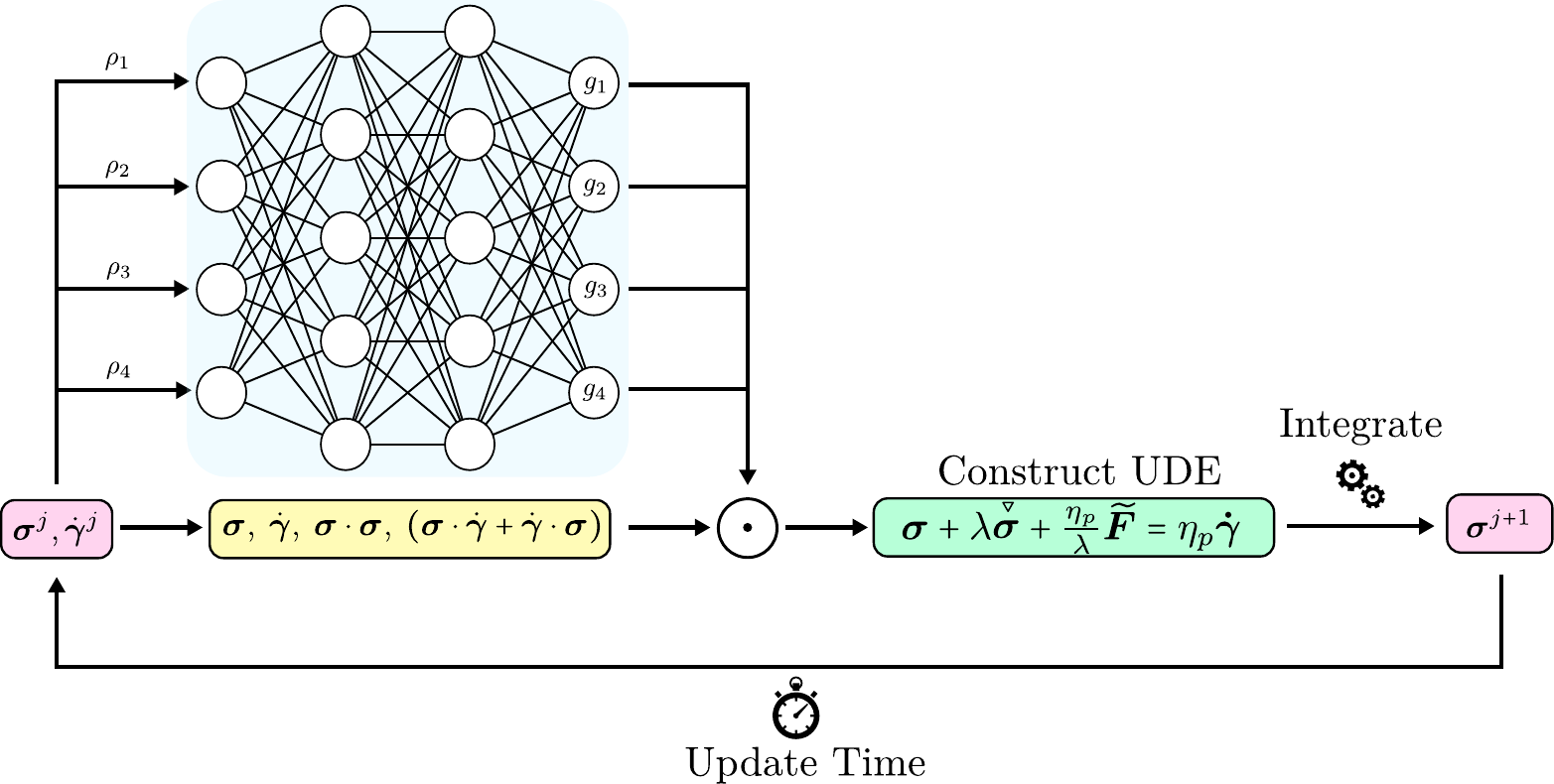}
    \caption{The integration sequence which generates the UDE stress responses employed in the calculation of the loss function. At a given time level $j$, the stress-strain-rate state is used to construct both the integrity basis tensors and the set of invariants $\rho_i$. The tensor basis neural network takes the invariants $\rho_i$ as inputs and outputs a set of coefficients $g_i$, which are then combined with the integrity basis to construct $\widetilde{\bm{F}}$ and thus the UDE for that time step. The UDE describing the stress evolution at time step $j$ is integrated to obtain the stress state at time $j+1$. The two hidden layers of the neural network each contain 32 neurons and the diagram is schematic only. The weights and biases of the TBNN are the same throughout the entire time horizon of a given integration.}
    \label{fig:1}
\end{figure}

Initially, the weights of the neural network are randomly assigned \cite{glorot2010understanding}, and the biases are initialised as zero. For each training epoch, the UDE is employed under exactly the same oscillatory shear protocols as was used to generate the training data (i.e. the same amplitudes and frequencies), producing a set of stress evolution signals directly analogous to the ground-truth stress signals. The signals generated by the UDE are compared with the corresponding ground-truth signals, and the neural network parameters are updated to reduce the error between them. When the training stage is completed, the TBNN has effectively learned a single set of weights and biases which is capable of producing a differential constitutive model which, when integrated across all of the deformation protocols, accurately reproduces the stress signals comprising the training data. The integration sequence of the UDE is elucidated in Figure \ref{fig:1} - at time level $j$, the current state of the fluid $\bm{\sigma}^j$ and $\dot{\bm{\gamma}}^j$ informs the TBNN inputs as well as the integrity basis tensors. The TBNN outputs and basis tensors are combined and the UDE is constructed and evolved in time, generating $\bm{\sigma}^{j+1}$, the stress tensor at time level $j+1$. It is important to emphasise that during each training epoch the same set of weights and biases is used to generate the stress responses for all of the training protocols; simultaneously, the outputs of the TBNN vary at each time step during the integration sequence, and a new inference must be generated for every time step.

Consistent with the typical TBNN approach, the inputs to the NN are the respective first invariants (i.e. the trace) of the 2D integrity basis tensors. We note at this stage that all of the inputs to the neural network, as well as the integrity basis tensors themselves, are made dimensionless according to the polymeric viscosity and relaxation time associated with the base model. 

The non-dimensional inputs to the neural network are thus given as in Eq. \ref{eq:10}, where $G_0 = \frac{\eta_p}{\lambda}$ is the elastic modulus associated with the base model:

\begin{equation}
\left\{
\begin{aligned}
\rho_1 &= \textnormal{tr}\!\left(\frac{\boldsymbol{\sigma}}{G_0}\right), \\[4pt]
\rho_2 &= \textnormal{tr}\!\left(\frac{\boldsymbol{\sigma} \cdot \boldsymbol{\sigma}}{{G_0}^2}\right), \\[4pt]
\rho_3 &= \textnormal{tr}\!\left(\lambda^2 \left(\boldsymbol{\dot{\gamma}} \cdot \dot{\boldsymbol{\gamma}}\right)\right), \\[4pt]
\rho_4 &= \textnormal{tr}\!\left(\frac{\lambda}{G_0} \left(\boldsymbol{\sigma} \cdot \dot{\boldsymbol{\gamma}}\right)\right)
\end{aligned}
\right.
\label{eq:10}
\end{equation}

\noindent Similarly, each neuron at the output layer is associated with a term belonging to the integrity basis:

\begin{equation}
\widetilde{\bm{F}}(\bm{\sigma}, \dot{\bm{\gamma}};\theta) = g_1 \left( \frac{\bm{\sigma}}{G_0} \right) + g_2 \left ( \lambda \dot{\bm{\gamma}} \right) + g_3 \left( \frac{\bm{\sigma} \cdot \bm{\sigma }}{G_0^2} \right) + g_4 \left( \frac {\lambda } {G_0} \left( \bm{\sigma } \cdot \dot{\bm{\gamma}} + \dot{\bm{\gamma}} \cdot \bm{\sigma} \right )   \right ) 
\label{eq:11}
\end{equation}

\noindent and therefore $g_1,\; g_2,\; g_3$ and $g_4$ can be thought of as the coefficients corresponding to terms in $\bm{\sigma}, \dot{\bm{\gamma}}, \left( \bm{\sigma} \cdot \bm{\sigma}\right), \left (  \bm{\sigma} \cdot \dot{\bm{\gamma}} + \dot{\bm{\gamma}} \cdot \bm{\sigma}  \right ) $ respectively.
It has already been observed that normalisation against material properties can improve the performance of the training step \cite{lennon2023scientific}, and it is generally well-established that the performance during training is improved when the numerical values at the input layer are $O(1)$ \cite{ioffe2015batch}.

The deviation relative to the ground-truth data is calculated based on the mean squared error (MSE):

\begin{equation}
\mathcal{L}_{\mathrm{MSE}} 
= \frac{1}{N_d} \sum_{b=1}^{N_d}
\left[
\frac{1}{T_b}
\sum_{t=1}^{T_b}
\frac{
\left(
\sigma_{12,t}^{(b)} - \hat{\sigma}_{12,t}^{(b)}
\right)^2
}{
\sigma_{12,\max}^{(b)}G_0
}
\right]
\label{eq:12}
\end{equation}

\noindent where $N_d$ is the number of datasets (i.e. oscillatory shear protocols) employed during training, $T_b$ is the total number of time steps used to discretise the entire temporal horizon of the flow; ${\sigma}_{12,t}^{(b)}$ and $ \hat{{\sigma}}_{12,t}^{(b)} $ represent the shear stress values for a given protocol and time step from the ground-truth and the UDE prediction at time $t$ ($0 \leq t \leq t_f$), respectively. Since the amplitudes of the stress signals span a wide range of magnitudes, the choice of normalisation determines how errors from each dataset are weighted within the loss function. Dividing the error associated with each dataset by $G_0^2$ yields an absolute mean-squared-error–type formulation. Under this normalisation, the loss can be dominated by large-amplitude stress responses, potentially biasing the training towards high-strain regimes at the expense of accurately capturing small-amplitude dynamics. Alternatively, normalisation by $\sigma_{12,\max}^2$, with $\sigma_{12,\max}^{(b)} = \displaystyle \max_{0 \le t \le t_f} \left| \sigma_{12,t}^{(b)} \right|$, produces a relative error measure in which all datasets contribute equally with respect to their stress amplitudes. While this prevents dominance by large-amplitude signals, it may over-emphasise low-stress regimes where absolute errors remain small and may fall within experimental or numerical noise. Here, we adopt a mixed normalisation based on both the maximum stress amplitude of each dataset and the elastic modulus of the base model. This choice balances the competing priorities of absolute and relative error formulations, mitigating excessive weighting of either high- or low-amplitude regimes, while rendering the loss dimensionless and anchored to a physically meaningful stress scale.

Unlike a Newtonian fluid, which only develops shear stresses under an imposed oscillatory shear flow, viscoelastic fluids also develop normal stresses \cite{hyun2011review}. We thus consider two cases for the training. In the majority of training cases, the neural network only sees the shear component (i.e. $\sigma_{12}$) of the stress tensor during training because this component is easier to measure experimentally. However, it is also possible to experimentally measure the first normal stress difference \cite{king2025self}, $N_1 = \sigma_{11}-\sigma_{22}$,  and so in some instances the neural network is trained on both the shear stress and the first normal stress difference during training. In this case, the $\mathcal{L}_{MSE}$ is given by:

\begin{equation}
\mathcal{L}_{\mathrm{MSE}} 
= \frac{1}{N_d} \sum_{b=1}^{N_d}
\left[
\frac{1}{T_b} 
\sum_{t=1}^{T_b} 
\left( 
\frac{
\left(
\sigma_{12,t}^{(b)} - \hat{\sigma}_{12,t}^{(b)}
\right)^2
}{
\sigma_{12,\max}^{(b)}G_0
}
+ 
\frac{
\left(
N_{1,t}^{(b)} - \hat{N}_{1,t}^{(b)}
\right)^2
}{
N_{1,\max}^{(b)}G_0
}
\right)
\right]
\label{eq:13}
\end{equation}

\noindent In all cases, an $\mathcal{L}_1$ regularisation penalty equal to $\kappa||\theta_{TBNN}||_1$ with weighting $\kappa = 0.001$ is applied, and we also enforce a cyclic penalty, $\mathcal{L}_{cyc}$ by penalising any difference in the values of the stress at exactly the same point in the last two cycles, with a weighting of $w_{cyc} = 10^{-6} $. The addition of this penalty was originally motivated by the observation that the trained neural network produces a stress signal that can ``drift" over time, especially in start-up shear flows where the stress is expected to reach a steady response. We find that the addition of a cyclic penalty rectifies the long-time drift and also helps to enforce physical symmetry more generally. Thus, the total loss function, $\mathcal{L}_T$, minimised during training is given by:

\begin{equation}
\mathcal{L}_{T} 
= \mathcal{L}_{MSE} + w_{cyc}\mathcal{L}_{cyc} + \kappa||\theta_{TBNN}||_1
\label{eq:14}
\end{equation}

In contrast with previous works, we choose not to employ a curriculum learning style ``warm-restart" approach which gradually introduces the synthetic datasets into the cumulative training data pool during training; instead, the neural network is exposed to all the training data immediately from the beginning of training. The staggered approach used previously was to alleviate issues related to the presence of an ``instability" during training, possibly leading to divergence. Instabilities and solvability problems during the training of neural ODEs are well-documented in the literature \cite{tuor_constrained_2020, thummerer_eigen-informed_2023}; however, in the present work we do not observe such instabilities provided that the strain amplitude of oscillatory shear remains within the moderately non-linear regime. In our experience, once the strain amplitude is sufficiently high, a curriculum learning approach is generally incapable of removing the instability and it is likely that more complex approaches will be required to obviate these difficulties.

In the present work, the neural network has two hidden layers with thirty-two neurons per hidden layer. For consistency with previous works \cite{lennon2023scientific}, the hyperbolic tangent function is selected as the non-linear activation function applied to the output of each neuron in the hidden layers; any activation function which is differentiable everywhere (e.g. softplus) is also equally suitable in principle. Both Adam \cite{kingma2014adam} and L-BFGS \cite{liu1989limited} optimisers are used to minimise the total loss function $\mathcal{L}_T$; the Adam optimiser is used first, followed by L-BFGS to attempt to find the global minimum and refine the loss further, which is a well-established approach commonly applied for PINNs. We also conducted training runs employing only an extensive Adam optimisation and found very little difference compared to the combined Adam and L-BFGS approach. When numerical integration is parameterised by the outputs of a neural network, a backpropagation-based optimiser must either have explicit access to the intermediate states of the integration in order to apply the chain rule and propagate gradients back through the numerical integration to the neural network, or alternatively solve a corresponding adjoint equation backwards in time to compute the sensitivities of the output variables with respect to the network parameters \cite{chen2018neural}. Here, we opt for the former approach by augmenting exactly the same numerical integrator used to generate the ground-truth stress signals to accommodate the neural network, thus granting the backpropagation algorithm full knowledge of the numerical integration sequence. We find that, for the networks considered in this study, this approach is not excessively demanding of memory at training time; consequently, one of the main motivations for adjoint-based sensitivity analysis (reduced memory usage) is not critical in our work.

\section{Deployment in OpenFOAM and rheoTool}
\label{sec:DeployCFD}

For all the numerical simulations performed here, we employ the finite-volume library rheoTool  \cite{pimenta_stabilization_2017} and consider laminar, isothermal, single-phase flows of incompressible fluids, with the conservation of mass and momentum equations given by: 

\begin{equation}
 \nabla \cdot \mathbf{u} = 0
\label{eq:15}
\end{equation}
\textcolor{black}{
\begin{equation}
\rho \frac{\partial \bm{u}}{\partial t} +  \rho \left( \bm{u} \cdot\nabla \right) \bm{u}
= - \nabla p+\nabla \cdot \boldsymbol{\sigma}  +\eta_s \nabla^2\bm{u}
\label{eq:16}
\end{equation}}where $\rho$ is the fluid density, $\bm{u}$ is the velocity vector, $p$ is the local pressure and $\bm{\sigma}$ is the viscoelastic extra-stress tensor given by either a ground-truth constitutive model or by a UDE. There is also a Newtonian solvent contribution to the stress, associated with the solvent viscosity $\eta_s$. The viscoelastic extra-stress evolution equation is given by:

\begin{equation} \boldsymbol{\sigma} + \lambda \stackrel{\kern0.0005em\smalltriangledown}{\boldsymbol{\sigma}} +   \boldsymbol{F} = \eta_p \dot{\bm{\bm{\gamma}}} 
\end{equation}
\label{eq:17}

\noindent where $\bm{F}$ is an additional non-linear term whose functional form is model dependent and given in Table 2. When the UDE is employed, then $\bm{F} = \frac{\eta_p}{\lambda}\widetilde{\bm{F}}(\bm{\sigma}, \dot{\bm{\gamma}}; \theta)$. The solvent viscosity ratio, $\beta$, is defined as the ratio of solvent viscosity to total zero-shear-rate viscosity: 

\begin{equation}
\beta = \frac{\eta_s}{\eta_0}
\end{equation}
\label{eq:18}

\noindent where $\eta_0 = \eta_p + \eta_s$.

\begin{figure}[htp]
    \centering
    \includegraphics[width=0.5\textwidth]{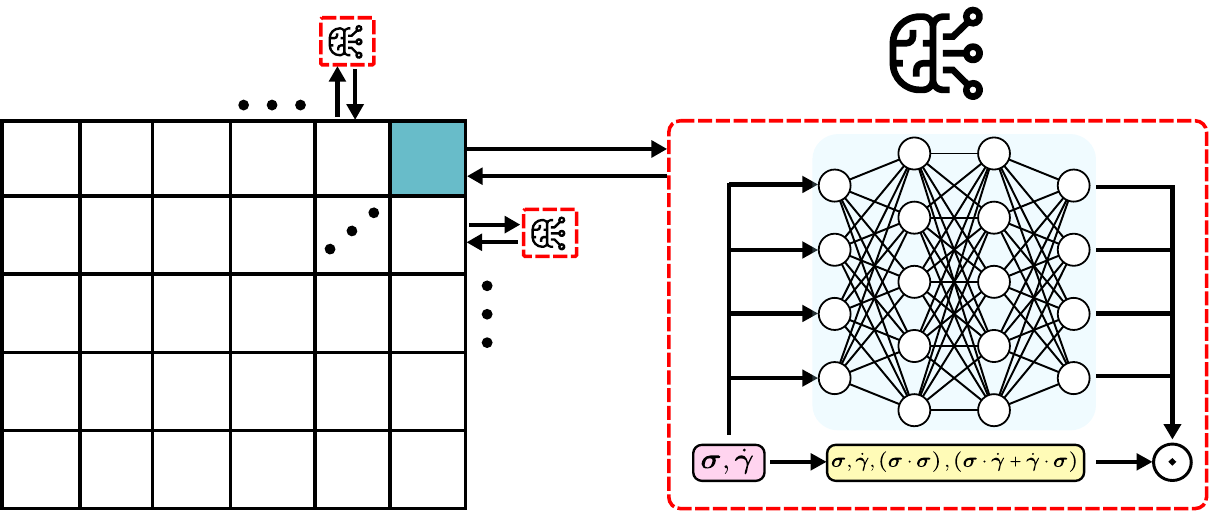}
    \caption{A schematic representation of the embedded TBNN in a finite-volume solver. The stress-strain-rate state for the cell highlighted in blue is fed as the inputs to a TBNN, which generates the correction term producing the constitutive model for that cell in the mesh. Each cell in the mesh undergoes the same procedure, calling a TBNN instance to generate the constitutive model describing the stress evolution for that cell.}
    \label{fig:2}
\end{figure}

The UDE framework comprises two main steps: evaluating the integrity basis tensor quantities and then performing the neural network inference. These are combined to produce the constitutive model. In general, the stress and shear rate vary in space and time, and it is therefore crucial that a CFD implementation accounts for the local and transient fluid state by evaluating the integrity basis tensors and the neural network inference separately for each cell in the mesh and at every time step as schematically shown in Figure \ref{fig:2}. This is also why an ``offline" approach (i.e. directly hard-coding the numerical result of a single neural network inference) which has been used to estimate the anisotropy tensor in turbulent flows \cite{ling_reynolds_2016} is unsuitable for the present work.

The UDE constitutive model itself is implemented as a model within the framework of rheoTool, and the neural network inference step is achieved by directly implementing the required series of matrix calculations and activation functions. The trained neural network parameters are obtained from the training pipeline and deployed through a series of auxiliary functions populating the matrix multiplications. The output of the forward pass is returned to the constitutive model where the tensor-basis combination is reconstructed and the model equation is formulated with the neural network contribution included in the source terms. In the present implementation, the inputs $\rho_1$ to $\rho_4$ are compiled as scalar fields and the neural network inference is performed on a field by field basis, as opposed to a cell-by-cell approach. Hence, the inference is vectorised over the fields so that only one NN call is performed per time step, and the overhead of calling the network does not grow significantly with mesh size, even though the number of arithmetic operations scales with the number of cells. As a consequence of the non-dimensionalisation of the inputs and the fact that the NN is embedded in the polymeric stress evolution equation, the base model parameters and solvent viscosity ratio may be varied in any numerical simulation where the UDE is deployed if the UCM model is selected as the base model. It is important to note that the solution algorithm is unmodified; the TBNN is used to generate the UDE describing the stress evolution, but the main solver loop is unmodified and solves the constitutive model in the usual manner as if it were defined via a classical closed form equation (cf. the pseudocode solver loop in Section 2.9 of \cite{pimenta_stabilization_2017}).

\section{Results} \label{sec:results} 

\subsection{Analysis of the learned representation of the TBNN
} \label{subsec:modouts}

After training, the behaviour learned by the neural network can be analysed by evolving the UDE in time under each oscillatory shear protocol used during training. For a given amplitude–frequency pair, each time step in the oscillatory shear cycle corresponds to a particular stress–strain-rate state. This state defines a set of invariants, $\rho_i$, which are supplied as inputs to the TBNN. The trained TBNN then predicts the corresponding coefficients, $g_i$, for that state. By recording the input invariants $\rho_i$ and the predicted coefficients $g_i$ at every time step across all training protocols, the learned coefficient mappings of the TBNN can be directly visualised. This provides insight into 
the bounds of the invariant space (specifically, the maximum and minimum values of each $\rho_i$) encountered during training and the behaviour of the TBNN within these bounds. When discussing the input-output relationships that the neural network has learned from the training step, reporting a single coefficient for each tensor basis is avoided because the coefficients for each tensor basis are not necessarily constant functions of the input space, and in fact are usually non-linear. 

\subsubsection{Giesekus and Johnson-Segalman Fluids}

\begin{figure}[htp]
    \centering
    \includegraphics[width=0.54\textwidth]{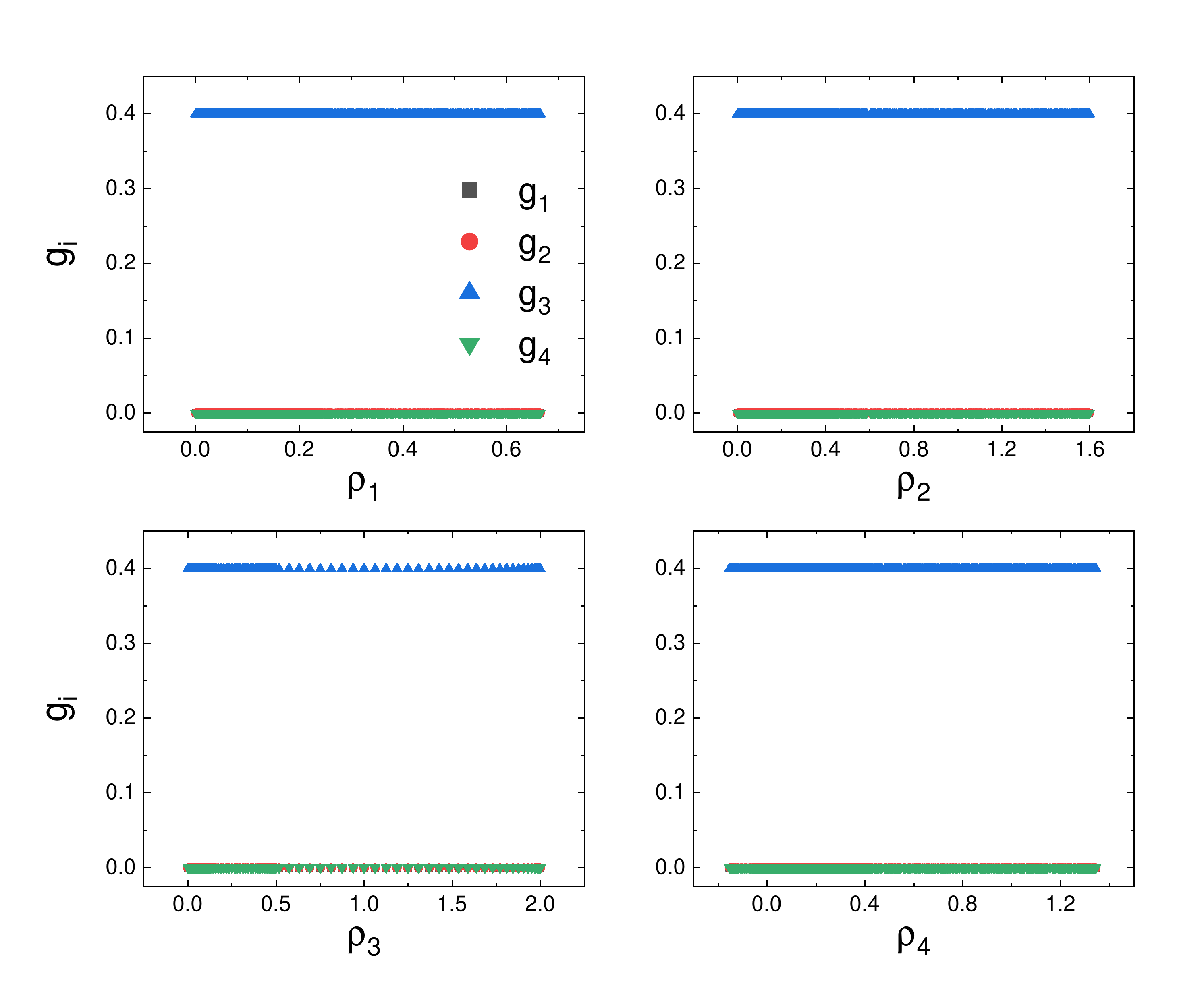}
    \caption{Variation of TBNN outputs $g_1$ to $g_4$ with each of the scalar invariants for a UDE trained on the Giesekus fluid ($\alpha = 0.4$). The points corresponding to $g_1$, $g_2$ overlap with $g_4$ visible in the graphs. }
   \label{fig:3}
\end{figure}

\begin{figure}[htp]
    \centering
    \includegraphics[width=0.63\textwidth]{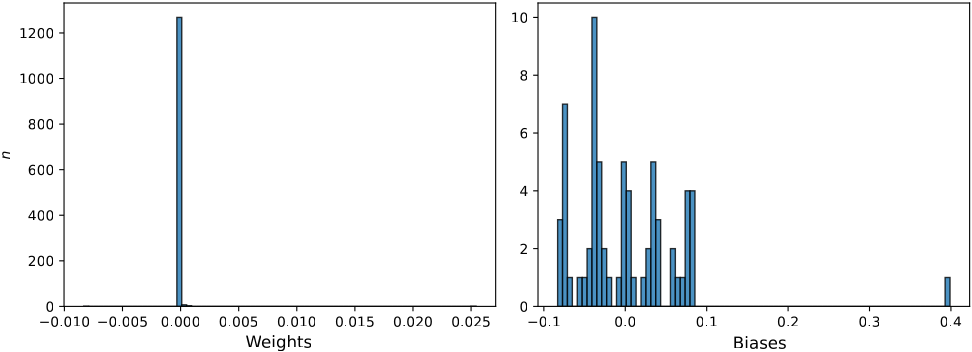}
    \caption{Histograms showing the distribution of the weights (left) and biases (right) of the neural network trained on a Giesekus fluid ($\alpha = 0.4$).}
    \label{fig:4}
\end{figure}

Focusing firstly on the Giesekus fluid, the neural network manages to accurately capture the analytical form of the constitutive model within the training envelope, with $g_3 \approx \alpha$ and $g_1, g_2, g_4 \approx 0$ across all of the input permutations (Figure \ref{fig:3}). Direct inspection of the weights and biases of the trained neural network for this case reveals that the vast majority of the weights are exactly zero, while the distribution of biases is around zero except for exactly one value associated with the output neuron for $g_3$, which is approximately equal to target Giesekus mobility parameter $\alpha$ ($g_3 = 0.399$ and $\alpha = 0.4$) (Figure \ref{fig:4}). It is worth recalling that the TBNN is able to isolate the actual value of $\alpha$, untethered to the polymeric viscosity and relaxation time of the ground-truth fluid data used during training, thanks to normalisation against material properties. In this case, the neural network has correctly learned to ignore the feature set at the input layer and simply produce the mobility parameter at the output layer, and from this perspective it is not altogether surprising that the network has recovered the analytical ground-truth equation.

Similarly, the TBNN trained on a Johnson-Segalman fluid (not shown) captures the analytical form of the constitutive model. In this case $g_4 \approx \frac{\xi}{2}$ and $g_1, g_2, g_3 \approx 0$ for all of the input configurations, and direct inspection of the weights and biases reveals an analogous result to the Giesekus model with the sole neural network parameter of significance being the relevant bias at the output layer. 
In these particular cases, the TBNN can easily and completely recover the correct form of the underlying constitutive equation when the target fluid exactly coincides with a constant-valued function of one of the tensors in the basis. Naturally, when the TBNN recovers the underlying rheology in this manner, it is able to accurately reproduce all non-zero components of the stress tensor in the LAOS protocols supplied during training, even when it is only exposed to one component (i.e. shear stress). As shall be observed throughout the remainder of the analysis, the recovery of the authentic ground-truth model results in almost perfect recovery of the ground truth model response, extending beyond LAOS and to all other tests. 

\begin{figure}[htp]
    \centering
    \includegraphics[width=1\textwidth]{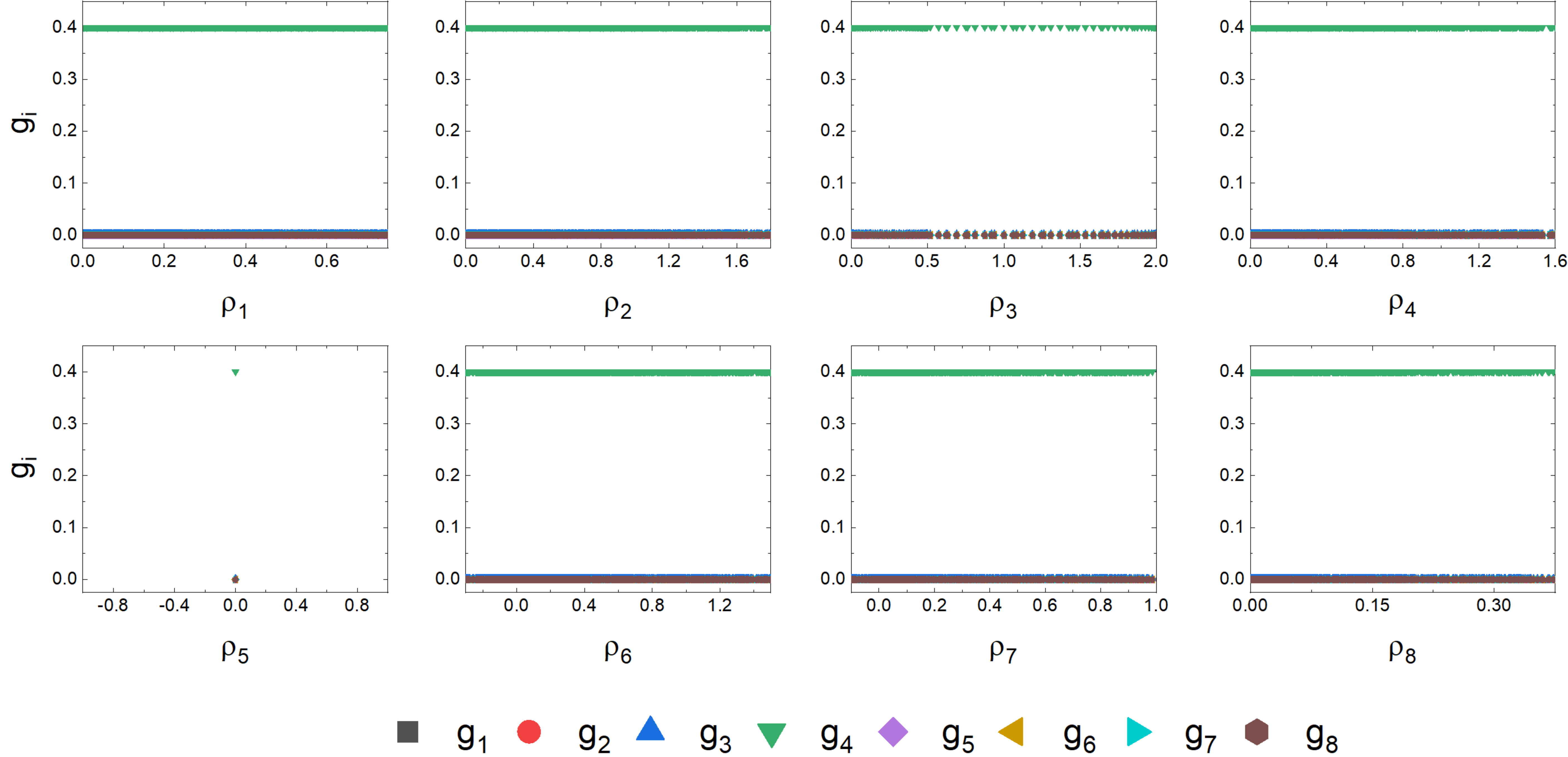}
    \caption{Variation of the 8 basis TBNN coefficient outputs $g_1$ to $g_8$ with each of the scalar invariants for a UDE trained on the Giesekus fluid ($\alpha = 0.4$).}
    \label{fig:5}
\end{figure}

\begin{figure}[htp]
    \centering
    \includegraphics[width=0.8\textwidth]{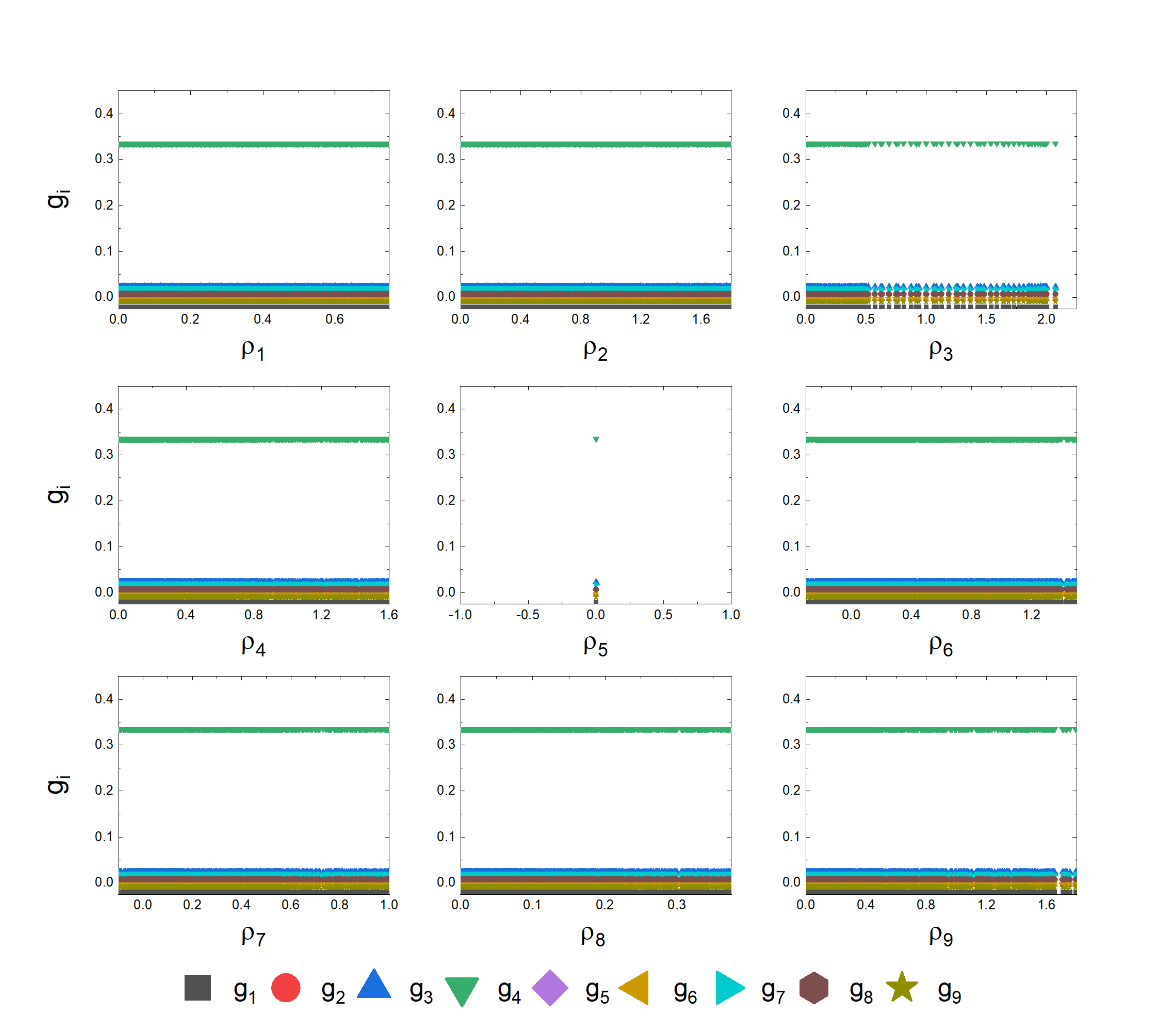}
    \caption{Variation of the 9 basis TBNN coefficient outputs $g_1$ to $g_9$ with each of the scalar invariants for a UDE trained on the Giesekus fluid ($\alpha = 0.4$).}
   \label{fig:6}
\end{figure}

We also examined the learned representations of the TBNNs trained on Giesekus fluids when the TBNN employs 8 basis tensors \cite{rodrigues2025finding} or 9 basis tensors \cite{lennon2023scientific}. Training these models is more challenging than with the proposed 4 basis model, since the numerical integration step is more susceptible to divergence and also because the presence of redundant basis tensors makes it more difficult for the TBNN to capture the parsimonious equation describing the underlying rheology. Considering the 8 basis model, the TBNN is able to essentially capture the authentic Giesekus model contribution (Figure \ref{fig:5}). However, many of our training runs with the 9 basis TBNN typically resulted in a dominant response in the Giesekus parameter, but with interference from other basis tensors (Figure \ref{fig:6}). In these cases, the additional 3D basis tensors are necessarily linearly dependent on the 2D basis tensors, and thus encode the same information. The presence of these additional redundant terms frustrates the ability of the TBNN to fully isolate and recover the parsimonious contribution to the Giesekus model. 

In any case, it is worth emphasising that the 4, 8 and 9 basis tensor model variants are in principle equally capable of learning to recover the authentic Giesekus model. However, in practice the 8 and 9 basis models require careful initialisation, training data selection and tuning of the training hyper-parameters to avoid numerical instabilities and recover the most relevant basis tensor contributions. In contrast, the 4 basis model is able to more reliably converge on the authentic underlying rheology in this case.

\subsubsection{Simplified, Linear and Exponential PTT Fluids}

Considering the network trained on a PTT fluid, the role of the neural network is more challenging. . 
Unlike the Giesekus model, it is not sufficient for the neural network to simply learn to produce one correct output parameter for any given set of inputs. Rather, a well-trained network must be able to meaningfully modulate the correct input parameter (i.e. the invariant corresponding to the trace of the stress) to produce a reasonable output. To accurately represent the correct analytical form, it must also recognise that all other inputs are irrelevant, and learn to correctly disregard them by constructing a set of weights that correctly mitigates their influence on the final output.

After training the neural network on an sPTT fluid with $\epsilon = 0.25$, we find that the neural network does not recover a constitutive model resembling the parsimonious ground-truth sPTT model (Figure \ref{fig:9}). Before analysing the outputs, it is worthwhile to point out the behaviour of the losses during training. For the networks trained on the Giesekus and Johnson-Segalman fluids, only the shear stress component, $\sigma_{12}$, was exposed during training. However, reiterating that the normal stress components $\sigma_{11}$ and $\sigma_{22}$ are not necessarily zero, a cross-component validation loss may also be defined by directly taking into account the errors in the unseen stress components. It is important to point out that a cross-component validation loss can only be calculated here due to the use of synthetic data where all of the components of stress are available; in a realistic experimental set-up, it is not possible to isolate the individual effects of stresses in this manner. Nevertheless, we introduce it here as it provides insight into the performance of the UDE and its capabilities. The cross-component validation loss, $\mathcal{L}_{CCV}$, is computed as:

\begin{equation}\label{eq:19}
\mathcal{L}_{\mathrm{CCV}} 
= \frac{1}{N_d} \sum_{b=1}^{N_d} 
\left[ 
\frac{1}{T_b} \sum_{t=1}^{T_b} 
\Big(
\frac{({\sigma}_{11,t}^{(b)} - \hat{\sigma}_{11,t}^{(b)})^2}{\sigma_{11,\textnormal{max}}^{(b)}G_0} 
+ \frac{({\sigma}_{12,t}^{(b)} - \hat{\sigma}_{12,t}^{(b)})^2}{\sigma_{12,\textnormal{max}}^{(b)}G_0} 
+ \frac{({\sigma}_{22,t}^{(b)} - \hat{\sigma}_{22,t}^{(b)})^2}{\sigma_{11,\textnormal{max}}^{(b)}G_0}
\Big) 
\right]
\end{equation}

\noindent where the error in $\sigma_{22}$ is normalised against the maximum value of $\sigma_{11}$ as a safeguard for cases where the response in $\sigma_{22}$ is very small or zero, and so employing it as a divisor would distort the loss value.

\begin{figure}[htp]
    \centering
    \includegraphics[width=0.75\textwidth]{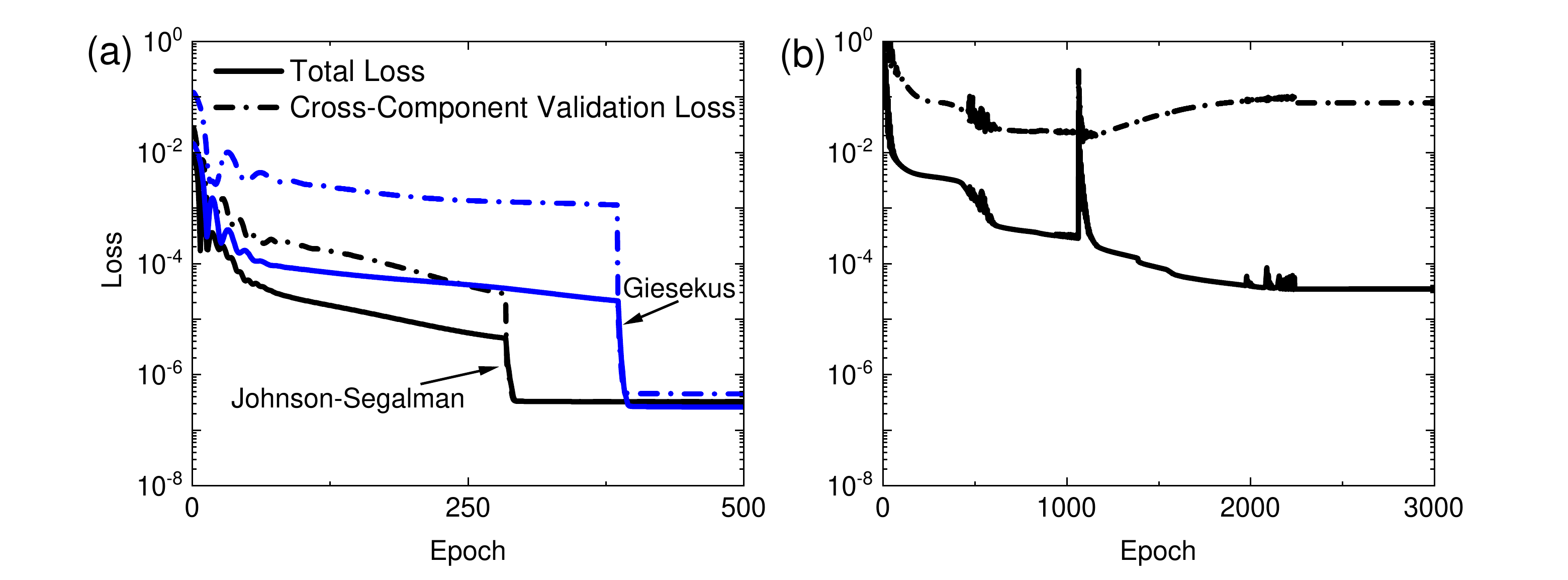}
    \caption{Typical evolution of the total loss used for training, $\mathcal{L}_T$, and the cross-component validation loss, $\mathcal{L}_{ccv}$ during the training procedure when the synthetic data is generated with: (a) Giesekus and Johnson-Segalman models; (b) The sPTT model ($\epsilon = 0.25$). The sudden decrease in loss in (a) is due to the introduction of the L-BFGS optimiser. }
    \label{fig:7}
\end{figure}

During the training of the networks on the Giesekus and Johnson-Segalman models, the cross-component validation loss drops to approximately the same value as the training loss (Figure \ref{fig:7}). This behaviour indicates that the error associated with the unseen stress components is negligible, which is consistent with the already established conclusion that the TBNN has recovered the authentic underlying rheology. However, during the training of the network on the sPTT fluid, the cross-component validation loss exhibits a plateau at several orders of magnitude above the training loss (in fact, it even increases whilst the training loss decreases). This behaviour in the losses is a characteristic of overfitting to the observed stress component, where a model adapts too closely to the training data at the expense of its ability to generalise to unseen data (Figure 1 of \cite{ying2019overview}). The clear implication in the present case is that the TBNN is unable to predict the unseen components of stress even in the same oscillatory shear protocols it is exposed to during training. This is a strong indication that the TBNN has been unable to recover the authentic ground-truth model in the same manner as the previously analysed cases where the network is trained on the Giesekus and Johnson-Segalman fluids. Instead, the TBNN learns a model which can accurately reproduce the training data but which does not accurately predict the unseen data, such as the normal stresses. The converged value of the training loss 
is also considerably higher (by almost two orders of magnitude) for the network trained on the sPTT fluid than for those trained with the Giesekus and Johnson-Segalman fluids. This is again attributable to the fact that the neural network does not recover the authentic form of the ground-truth constitutive model for the sPTT fluid, and in fact this behaviour is observed in all cases other than the Giesekus and Johnson-Segalman fluids, which are in this respect exceptions.

\begin{figure}[htp]
    \centering
    \includegraphics[width=1\textwidth]{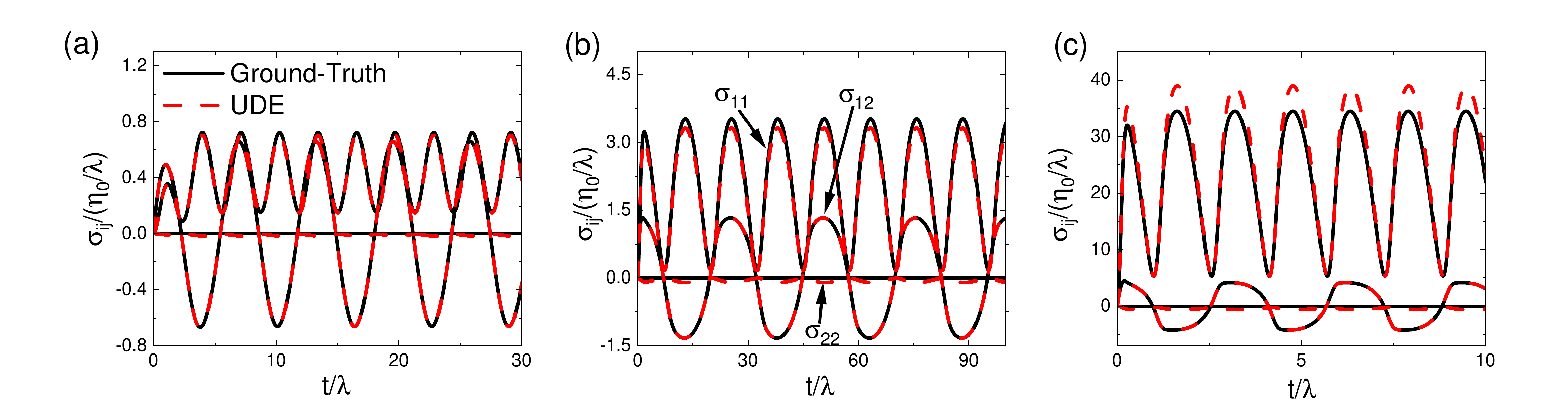}
    \caption{Comparison of the stress response of the trained UDE and corresponding ground-truth fluid (sPTT $\epsilon = 0.25$) for oscillatory flow with: (a) $Wi = 1, De = 1$; (b) $Wi = 2.5, De = 0.25$; (c) $Wi = 40, De = 2$.}
    \label{fig:8}
\end{figure}

\begin{figure}[htp]
    \centering
    \includegraphics[width=0.4\textwidth]{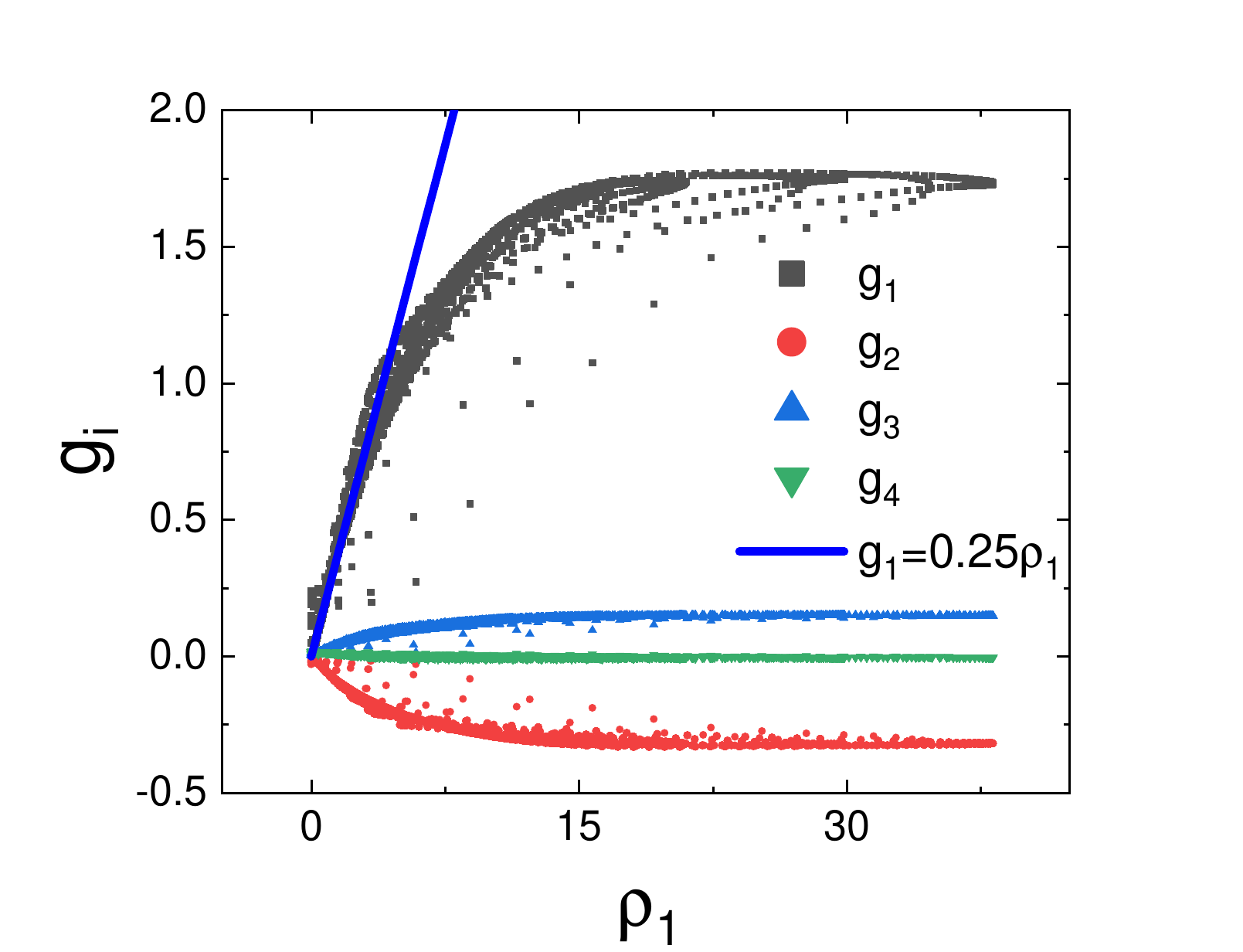}
    \caption{Variation of neural network coefficient outputs $g_i$ with the scalar invariant $\rho_1$ for a UDE trained on an sPTT fluid ($\epsilon = 0.25$).}
    \label{fig:9}
\end{figure}

By examining the UDE response to the LAOS protocols to which the UDE was exposed during training, a behaviour consistent with a larger cross-component validation loss can be directly observed as errors manifest in the prediction of the normal stress signals (i.e. $\sigma_{11}$ and $\sigma_{22}$) (Figure \ref{fig:8}). Indeed, while the UDE accurately reproduces the shear stress response, which has been seen during training, the signal of $\sigma_{11}$ is consistently misrepresented. The emergent pattern is generally an over- or under-prediction of its magnitude in the neighbourhood of the peaks. We observe that this error, also described by Rodrigues et al. \cite{rodrigues2025finding}, varies in magnitude depending on the conditions of the given LAOS protocol; in some cases it is almost negligible although in the most extreme cases (e.g. Figure \ref{fig:12}) the error exceeds 25\%. However, the magnitude of the error is less significant than its mere existence, as this alone provides further direct evidence that the UDE is not capable of fully capturing the authentic underlying rheology, which may have downstream implications in deployment to CFD calculations.

As shown in Table 2, the analytical form of the target function for the sPTT fluid is $\frac{\epsilon \lambda }{\eta_p}\textnormal{tr}(\bm{\sigma})\bm{\sigma} $. A neural network which recovers the authentic ground-truth model would therefore produce a linear variation in $g_1$, the coefficient multiplying $\bm{\sigma}$, with the gradient controlled by the value of $\epsilon$ used in the synthetic data employed during training. Although the neural network approximately captures this relationship for low values of $\rho_1$, the gradient in $g_1$ begins to flatten out after $\rho_1 \sim 10$, and becomes almost completely constant for $\rho_1 \geq 20$ (as shown in Figure \ref{fig:9}. Clearly, the learned responses of $g_2$ and $g_3$ (corresponding to coefficients in $\bm{\dot{\gamma}}$ and $\boldsymbol{\sigma} \cdot \boldsymbol{\sigma}$ respectively) have a compensating effect, with $g_2$ decreasing and $g_3$ increasing with $\rho_1$. Evidently, this combination is able to produce a similar response to the sPTT fluid for the shear component of stress in oscillatory shear flows, again illustrating model identifiability issues highlighted by John et al. \cite{john_machine_2024} when the training data exclusively comprises oscillatory shear data.  The output response plot of Figure \ref{fig:9} represents a projection of the full learned invariant space since the remaining invariants (i.e. $\rho_2$, $\rho_3$ and $\rho_4$) are not held fixed. As a result, points with the same value of $\rho_1$ can correspond to different invariant combinations. Notably, the distribution of $g_1$ values under these varying conditions is relatively narrow suggesting that the neural network has generally learned the importance of $\textnormal{tr}(\bm{\sigma})$ in determining the rheological behaviour of the fluid, and is less influenced by other input feature permutations so long as $g_1$ is constant.

In general, the learned TBNN representations of all of the UDEs broadly reflect the same behaviour, which is complex but ultimately not resembling the parsimonious form of the constitutive model originally employed to generate the data.

\begin{figure}[htp]
    \centering
    \includegraphics[width=0.7\textwidth]{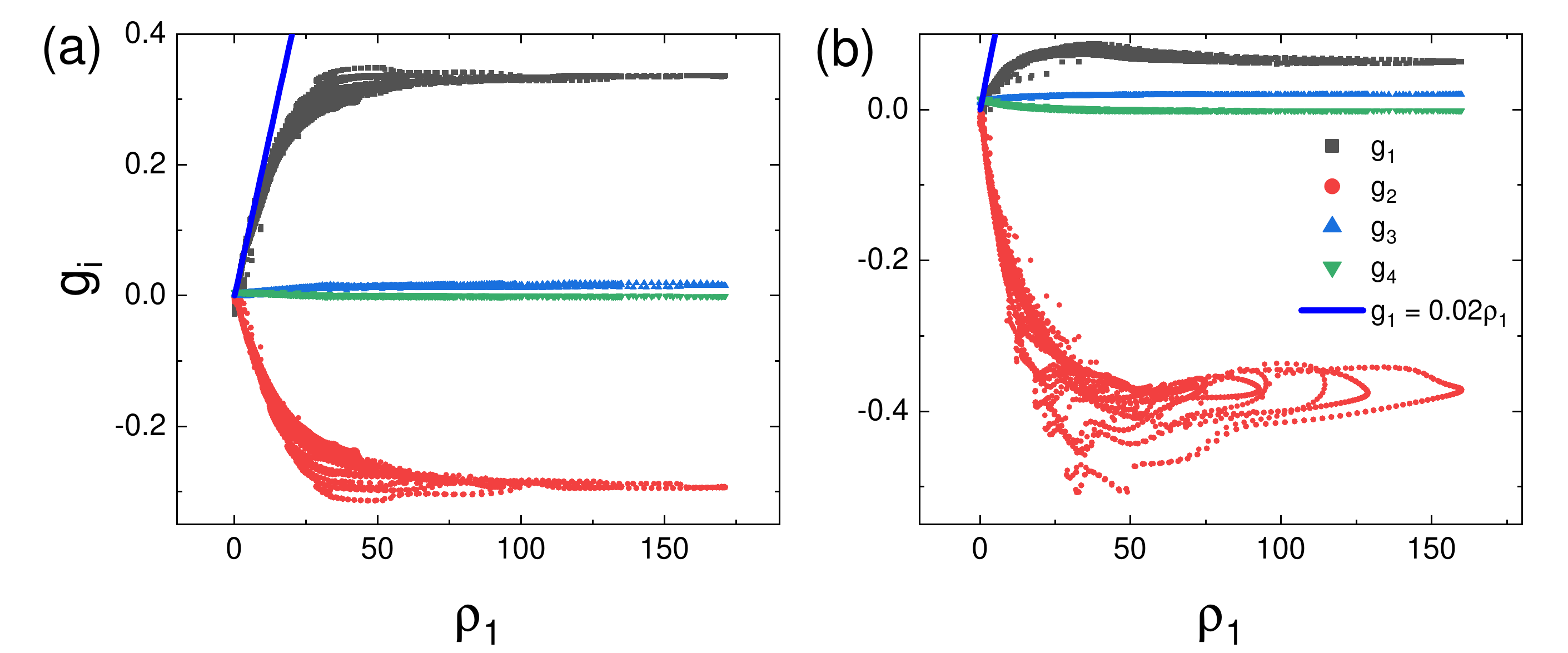}
    \caption{Variation of neural network coefficient outputs $g_i$ with the scalar invariant $\rho_1$ for a UDE trained on an sPTT fluid ($\epsilon = 0.02, \xi=0$). (a) The TBNN is exposed to $\sigma_{12}$ only during training; (b) The TBNN is exposed to $\sigma_{12}$ and $N_1$ during training.}
    \label{fig:10}
\end{figure}

\subsubsection{Effect of Exposure to $N_1$}

A natural question prompted by the larger cross-component validation loss and the observed error in the normal stresses is whether exposure to additional 
stress components helps the neural network to learn relationships which more closely resemble the authentic ground-truth model. Given that measurements of $N_1$ in oscillatory shear, though challenging, are feasible \cite{king2025self}, we now train the neural network using exactly the same set of protocols, but exposing not only the shear stress component but also the first normal stress difference. For this analysis, we consider the sPTT model and ePTT model with $\epsilon = 0.02$ and $\epsilon = 0.25$ respectively.

\begin{figure}[htp]
    \centering
    \includegraphics[width=0.7\textwidth]{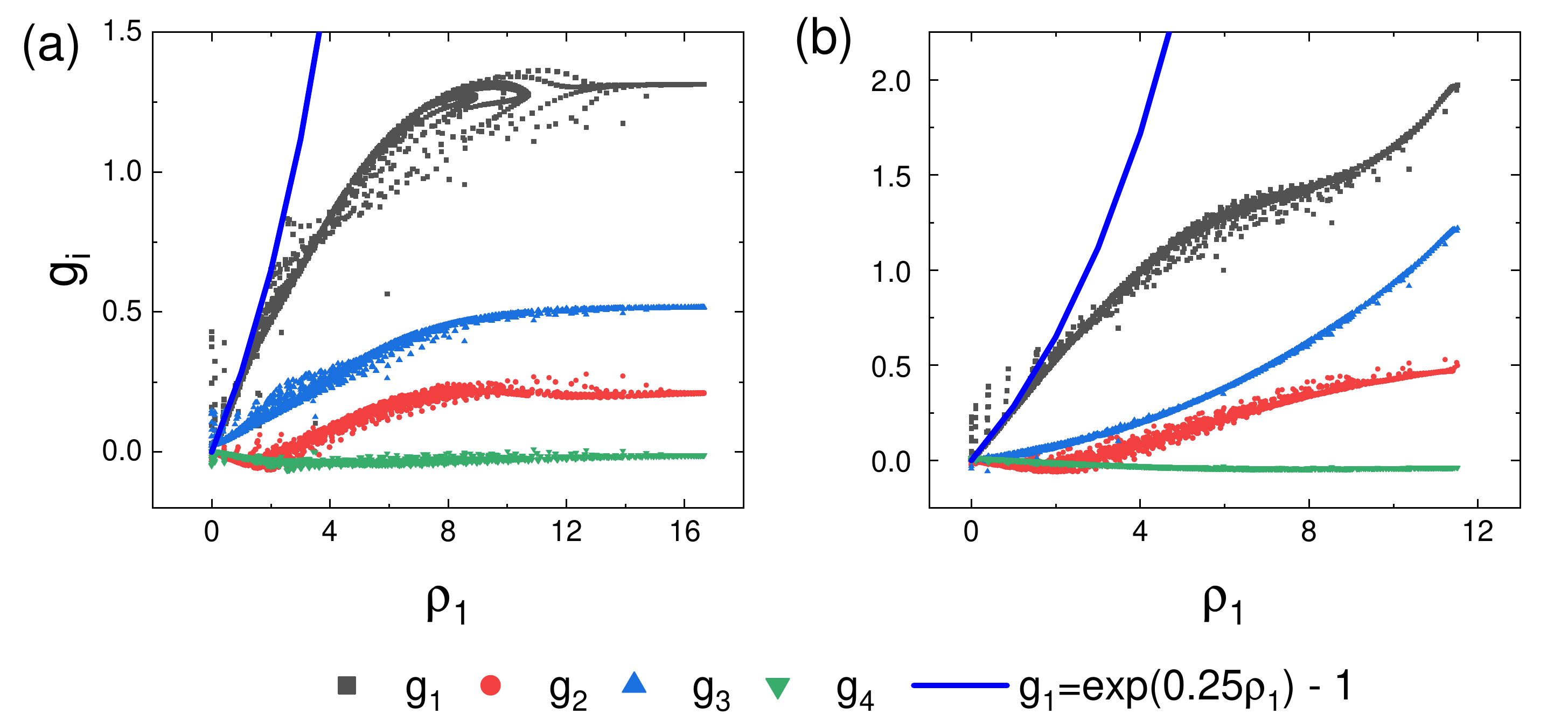}
    \caption{Variation of neural network coefficient outputs $g_i$ with the scalar invariant $\rho_1$ for a UDE trained on an ePTT fluid ($\epsilon = 0.25, \xi=0$). (a) The TBNN is exposed to $\sigma_{12}$ only during training; (b) The TBNN is exposed to $\sigma_{12}$ and $N_1$ during training.}
    \label{fig:11}
\end{figure}

\begin{figure}[htp]
    \centering
    \includegraphics[width=0.75\textwidth]{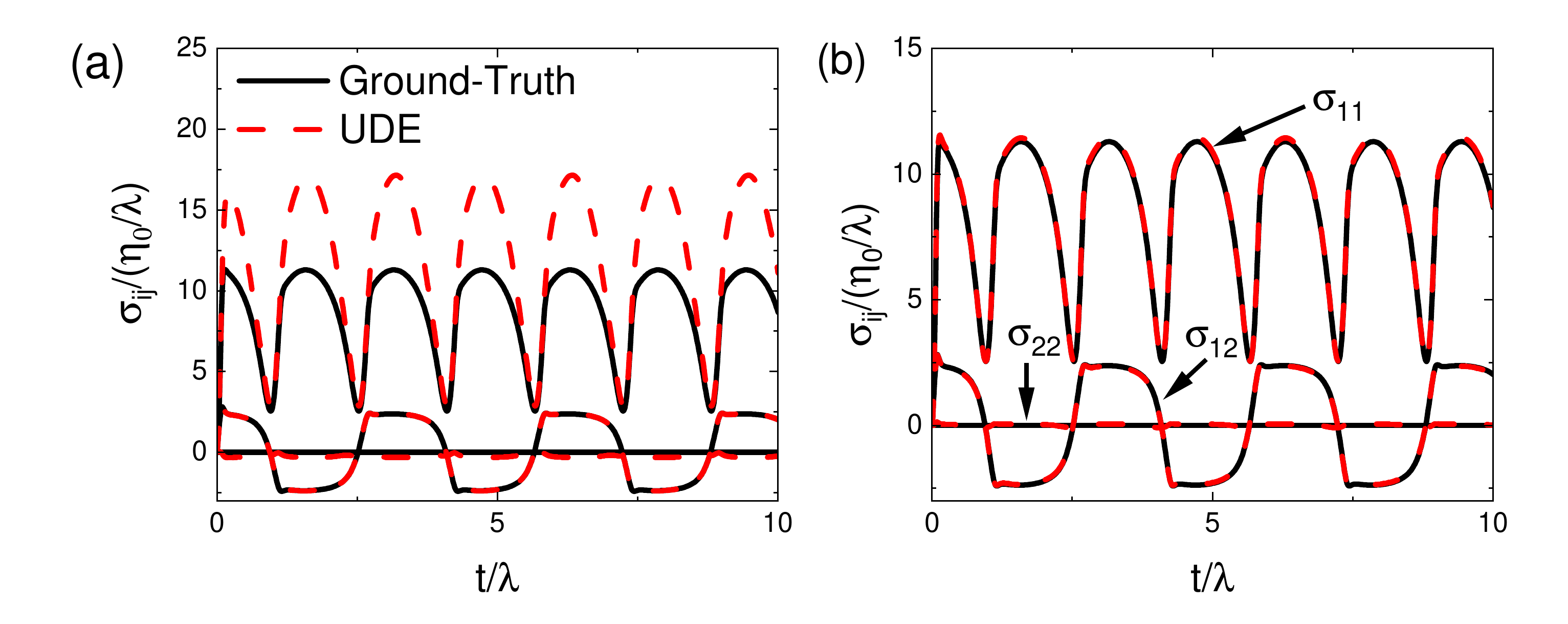}
    \caption{Comparison of the stress responses of the trained UDEs and the corresponding ground-truth fluid (ePTT $\epsilon = 0.25, \xi = 0$) for oscillatory flow with $(Wi, De) = (40, 2)$, which is directly sampled during training. (a) The UDE is exposed to the shear stress component only during training. (b) The UDE is exposed to both the shear stress and the first normal stress difference  during training.}
    \label{fig:12}
\end{figure}

As expected, including $N_1$ in the training loss and thus exposing it to the TBNN during training has the direct effect of ensuring that the trained UDE is able to almost perfectly replicate all three non-zero stress component signals in the LAOS protocols comprising the training data (Figure \ref{fig:12}). However, it also significantly changes the learned TBNN representation. Considering firstly the sPTT model (Figure \ref{fig:10}), it is interesting to note that introducing the first normal stress difference in training does not lead the TBNN to learn a model which more closely resembles the form of the ground-truth constitutive model. In fact, the TBNN learns a new behaviour for $g_1$, while the output associated with $g_2$ becomes more dominant. The behaviour in $g_3$, which imparts a non-zero second normal stress difference, is almost unchanged. The situation is similar considering the TBNN trained on ePTT fluids (Figure \ref{fig:11}): the TBNN learns a significantly different behaviour but not one that necessarily resembles the form of the ground-truth more closely. 

Although this now clearly establishes that the TBNN does not
reproduce the ground-truth constitutive equation
with either of the given training conditions, it remains impossible to deduce whether
generalisation to unseen flows is possible (and what extent) simply from observing the learned representation of the TBNN alone. Therefore, we seek to ascertain differences via direct deployment of the models in rheometric and complex flows and comparing their behaviour to the training fluids.

\subsection{Material Properties} \label{sec:matprops}

In this section, we assess the capabilities of the UDEs trained 
entirely using oscillatory shear flow data to reproduce key material properties, such as shear and extensional viscosity. The steady shear and extensional viscosities are obtained by simulating a start-up shear or planar extensional flow and considering the resulting stress once a steady state is reached.

It is worth noting that rheoTestFoam, a well-documented application available within rheoTool \cite{pimenta_stabilization_2017}, can also be used to obtain shear and extensional viscosity curves. 
The use of rheoTestFoam for determining material properties simplifies the workflow as it is fully contained within the OpenFOAM ecosystem after importing trained weights and biases. This approach was validated against the results of the integrator used during training. It is worthwhile to point out that the calculation of the start-up stress evolution at high values of $De$ becomes challenging from a numerical point of view. At these values of $De$, the ODE describing the stress evolution becomes stiff, and because the temporal derivative is generated by the TBNN which is extrapolated well beyond its training envelope, the severity of the stiffness is typically greater than in the ODE describing the stress evolution of the ground-truth constitutive model. In our experience, rheoTestFoam generally exhibits improved stability compared to our integration sequence which is fully explicit, although in many cases it was very challenging to evolve the stresses at high $De$ to obtain the full profile of material properties. In every case, the UDE performance is evaluated with material properties $\eta_0 = 1\;\textnormal{Pa} \cdot \textnormal{s}, \beta = 1/9$ and $\lambda = 1\; \textnormal{s}$, although in principle any values can be used due to the TBNN normalisation procedure. 

\begin{figure}[htp]
    \centering
    \includegraphics[width=1\textwidth]{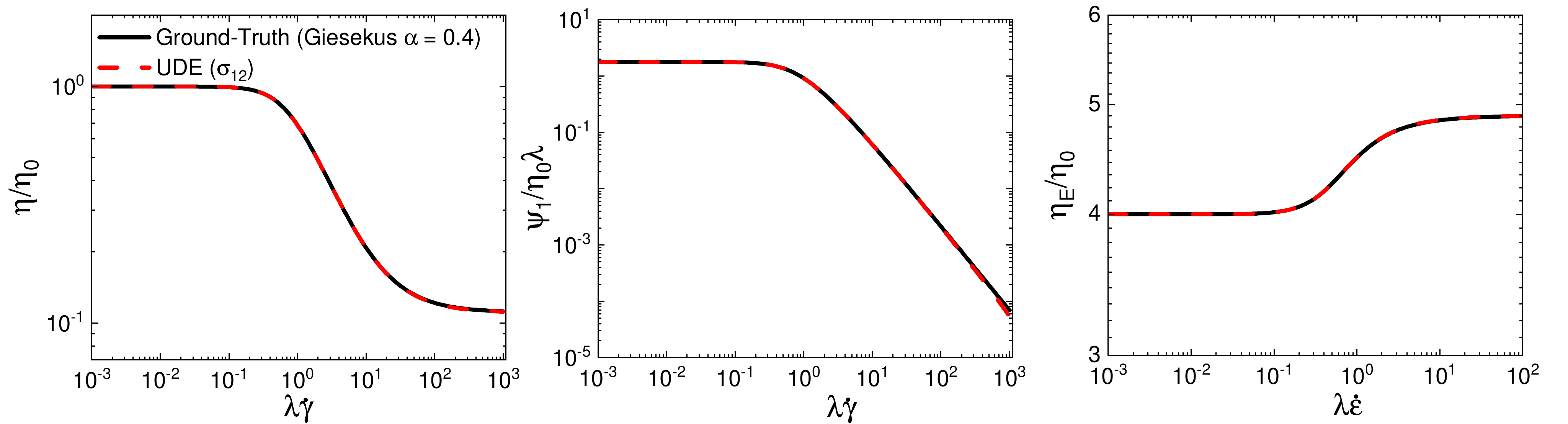}
    \caption{Comparison between the trained UDE and the corresponding Giesekus fluid ($\alpha = 0.4, \eta_0 = 1\; \textnormal{Pa}\cdot \textnormal{s, }\lambda = 1\;\textnormal{s, } \beta = 1/9 $) material properties: steady shear viscosity ($\eta/ \eta_0$) and  first normal stress difference coefficient ($\psi_1 / \eta_0 \lambda$) obtained in shear and  planar extensional viscosity ($\eta_E/ \eta_0$). }
    \label{fig:13}
\end{figure}

Both Lennon et al. \cite{lennon2023scientific} and Rodrigues et al. \cite{rodrigues2025finding} demonstrate that the UDE trained on the Giesekus fluid is able to correctly predict the transient response in start-up shear. Here, we observe that not only the steady shear viscosity curve but also the first normal stress difference coefficient is almost perfectly reproduced by the trained UDE, even for values of $De$ far beyond those encountered in training.      This is a natural consequence of the internal structure of the trained TBNN. Since the neural network effectively multiplies the inputs by weights close to zero at the input layer, large numerical inputs (even those which extrapolate very far beyond the invariant space exposed during training) do not destabilise the UDE. For the same reason, the UDE is also able to accurately reproduce the correct extensional viscosity (Figure \ref{fig:13}). This is a considerably more challenging test given that the UDE is not exposed to any kind of extensional flow during training, though in this case we see further confirmation that the UDE has indeed completely recovered the underlying constitutive model.

\begin{figure}[htp]
    \centering
    \includegraphics[width=1\textwidth]{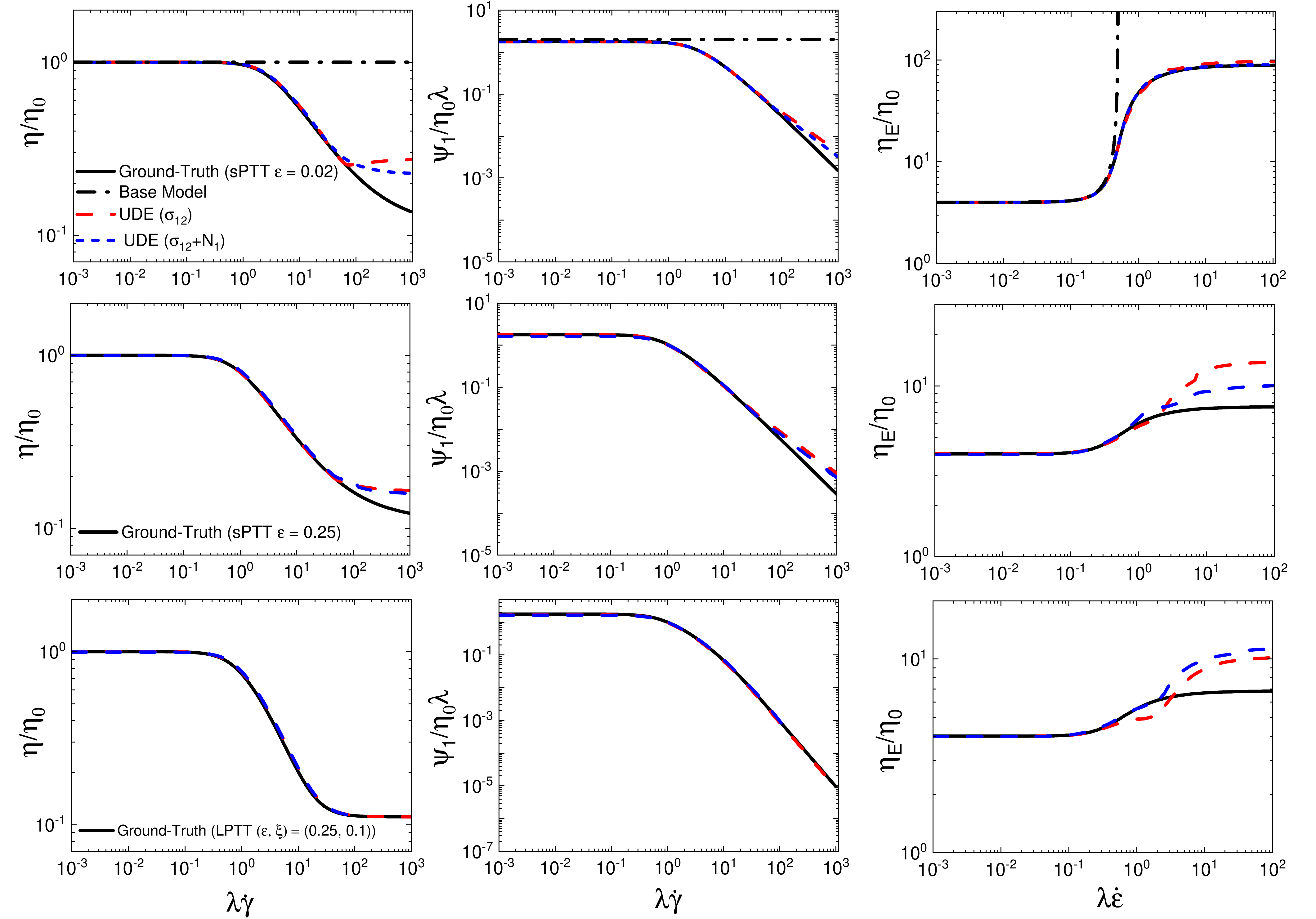}
    \caption{Material properties for the trained UDEs and the corresponding ground truth, considering sPTT ($\epsilon = 0.02, 0.25$) and LPTT fluids ($(\varepsilon, \xi) = (0.25, 0.1)$). For comparison, the base model without a TBNN correction is also shown. In all cases, we consider ($\eta_0 = 1\; \textnormal{Pa}\cdot \textnormal{s, }\lambda = 1\;\textnormal{s, }\beta = 1/9)$.}
    \label{fig:14}
\end{figure}

\begin{figure}[htp]
    \centering
    \includegraphics[width=1\textwidth]{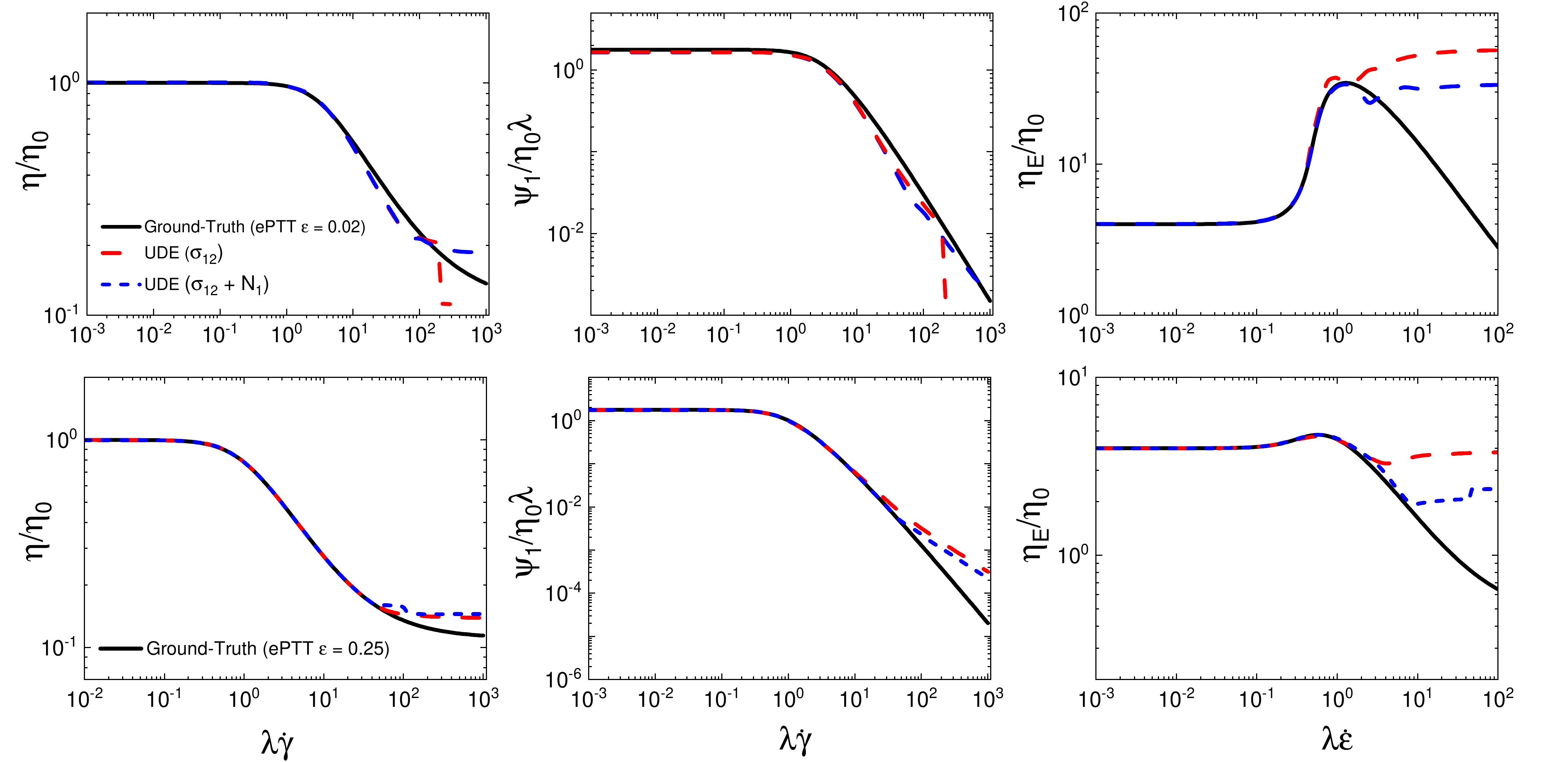}
    \caption{Material properties for the trained UDEs and the corresponding ground truth, considering ePTT fluids ($\varepsilon = 0.02, 0.25, \xi=0$). In all cases, we consider ($\eta_0 = 1\; \textnormal{Pa}\cdot \textnormal{s, }\lambda = 1\;\textnormal{s, }\beta = 1/9)$.  }
    \label{fig:15}
\end{figure}

Moving to the other cases where it has been determined that the TBNN does not capture the genuine underlying rheological model, the shear-dependent properties are, however, in almost every case accurately reproduced by the UDEs when the Deborah number is below the maximum observed during training (cf. Figures \ref{fig:14} and \ref{fig:15} for the linear/simplified and exponential PTT, respectively). Beyond the maximum Deborah number shown during training, which in every case here corresponds to $De = 40$, the shear-dependent properties (normalised shear viscosity, $\eta/\eta_0$, and first normal stress difference coefficients, $\psi_1 / \eta_0 \lambda$) 
begin to deviate from the ground-truth, with the shear viscosity exhibiting an early plateau for the cases trained on the sPTT and ePTT, but predicting accurately the whole viscosity curve for the case trained on the LPTT ($(\varepsilon, \xi) = (0.25, 0.1)$). Considering the extensional viscosity of the UDEs, we observe that in every case the correct extensional viscosity is reproduced in the limit of zero strain rate, corresponding to the Trouton ratio of $Tr = 4$ as expected for planar extension. Note that we choose to plot $\eta_E/\eta_0$, the ratio of extensional viscosity to zero-shear viscosity, rather than the Trouton ratio $\eta_E/\eta$, as is usual, to avoid introducing erroneous shear viscosity predictions for the high range of shear rates. Further, in every case the UDE is able to reproduce the onset of strain-hardening, capturing both the correct conditions of onset as well as shape. However, the accuracy with which the UDEs reproduce the entire extensional viscosity profile depends on the fluid on which they were trained. Considering the sPTT fluid, the entire extensional viscosity is reproduced with minimal error for the case with $\epsilon = 0.02$ (shown in the first row of Figure \ref{fig:14}), and exposure to $N_1$ during training seems to improve the UDE prediction to be almost indistinguishable from the ground-truth. However, for a large $\epsilon = 0.25$ (in both the sPTT and the LPTT 
cases shown in the second and third rows of Figure \ref{fig:14}, respectively), we see an over-prediction in the extensional viscosity for $\lambda \dot{\epsilon} > 2$. The reduced accuracy of the UDEs with increasing $\epsilon$ can be understood by examining the role of the TBNN as a corrective term: as $\epsilon$ increases, the required TBNN correction also increases in magnitude, and thus the TBNN must supply corrections which are both larger in magnitude and increasingly precise in order to recover the genuine sPTT material response. Considering the ePTT fluid (Figure \ref{fig:15}), the situation is similar. Modelling the network destruction function as exponential leads to a non-monotonic extensional viscosity profile reflecting the ``network rupture" effect observed in some fluids \cite{phan1978nonlinear}. Again, the accuracy of the UDE predictions varies with $\epsilon$, although interestingly the predictions in this case are better for $\epsilon = 0.25$, with the UDE
correctly capturing the overshoot in extensional viscosity as well as the onset of extensional thinning for this case.
Exposing the TBNN to $N_1$ during training grants the UDEs improved capabilities for capturing the extensional viscosity curve over a wider range of conditions (up to $\lambda \dot{\epsilon} = 8$ for the UDE exposed to $\sigma_{12}$ and $N_1$ compared with $\lambda \dot{\epsilon} \approx2$ for the UDE exposed to $\sigma_{12}$ alone). Indeed, for all the fluids examined here, exposure to $N_1$ during training improves the predictive capability in extension. 
The fact that the UDE exposed solely to shear stress is able to correctly capture a portion of the extensional viscosity profiles is an encouraging result.

\subsection{Validation and Numerical Stability in CFD Simulations} \label{subsec:cfdeployment}

The frame-invariant neural network architecture was also fully leveraged by deploying it in full CFD simulations using OpenFoam V9 with rheoTool \cite{pimenta_stabilization_2017}. Before proceeding to the test cases, where UDEs trained on fluids with different rheological properties are deployed in more complex flow configurations, such as the contraction (planar and fully 3D square-square) and cross-slot flows, our UDE implementation in rheoTool is validated. This is done by comparing LAOS simulations against the responses observed during training, followed by an assessment of numerical stability using a 2D slit flow as a benchmark case.  

\begin{figure}[htp]
    \centering
    \includegraphics[width=0.5\textwidth]{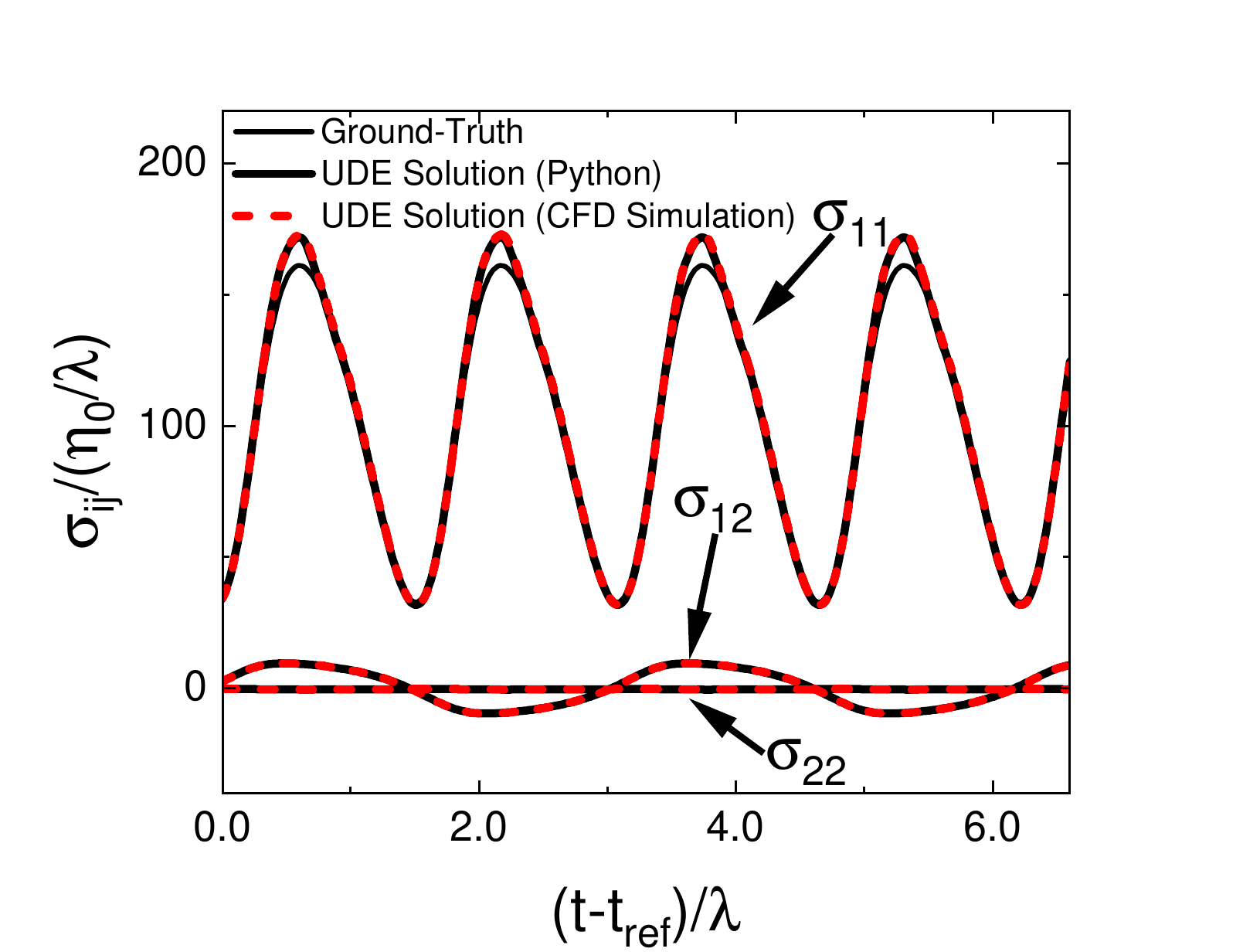}
    \caption{Comparison between the solution of a LAOS protocol at $(Wi, De) = (40, 2)$ using the simplified Python training pipeline, for a UDE trained on an sPTT fluid with $\epsilon = 0.02$, and the full CFD simulation of the same flow employing the UDE implementation in rheoTool. Data is presented after steady oscillations are reached, with a reference time $t_{ref}/\lambda=5.9$.}
    \label{fig:16}
\end{figure}

In order to establish the validity of our trained UDE constitutive model implementation, we first deploy the UDE in a LAOS simulation in rheoTool and compare the resulting stress evolution against the same evolution obtained using the integration scheme employed during training. During training, the stress evolution of the UDE is obtained via numerical solution of the simplified set of constitutive equation only, while the CFD solution solves the mass, momentum and stresses equations without simplification. The example shown here (Figure \ref{fig:16}) corresponds to the UDE trained on the sPTT model with $\epsilon = 0.02$ and exposed only to $\sigma_{12}$ during training, for
$(Wi, De) = (40, 2)$, which also coincides with one of the protocols encountered during training (Figure \ref{fig:16}). The CFD implementation exactly replicates the response observed in the training pipeline. This is true even when the trained UDE does not perfectly replicate all of the stress signals: in the case shown in Figure \ref{fig:16}, the peaks in $\sigma_{11}$ are over-predicted by the UDE in the solution generated via Python and also in the CFD solution obtained in OpenFoam compared to the ground-truth solution, which demonstrates that CFD is correctly inheriting the expected UDE response.

We considered the 2D slit flow as a test case for assessing numerical stability, and deploy the TBNNs employing nine, eight and four basis tensors in CFD calculations under identical conditions. All three UDEs are trained with exactly the same training data generated using the Giesekus model ($\alpha = 0.4$).

Considering the 9 basis model \cite{lennon2023scientific}, we find that the integrity basis tensor in $\bm{\sigma} \cdot \bm{\sigma} \cdot \dot{\bm{\gamma}} \cdot \dot{\bm{\gamma}} + \dot{\bm{\gamma}} \cdot\dot{\bm{\gamma} } \cdot\ \bm{\sigma} \cdot \bm{\sigma}$ is strongly associated with numerical instability. It is possible to directly link this term to the observed instability by artificially populating the biases of the neural network at the output layer and analysing their impact on numerical stability. For instance, by setting the values of all weights and biases in the neural network to zero, the TBNN contribution vanishes and the base UCM/Oldroyd-B model remains. By setting the value of the bias in the neuron associated with the output informing the coefficient of the basis tensor $\bm{\sigma} \cdot \bm{\sigma}$ to $0.4$ and keeping all parameters zero, the exact analytical form of the Giesekus model is recovered, with the same numerical behaviour as the existing Giesekus model implementation within rheoTool.

Simple noise in other channels may then be emulated by setting the biases of other neurons in the output layer to small values. By setting the value of the bias in the neuron associated with $g_9$ (the coefficient of the basis tensor $\bm{\sigma} \cdot \bm{\sigma} \cdot \dot{\bm{\gamma}} \cdot \dot{\bm{\gamma}} + \dot{\bm{\gamma}} \cdot\dot{\bm{\gamma} } \cdot\ \bm{\sigma} \cdot \bm{\sigma}$) to $0.001$, which corresponds to the ground-truth Giesekus model with a small amount of noise in the coefficient $g_9$, a numerical instability immediately manifests which makes it impossible to obtain a converged solution for the 2D slit flow despite the fact that this instability is not present for the same flow conditions in the ground-truth model or its artificial emulation. 

It is important to recognise that the imposition of a plug flow at the inlet of a 2D slit flow results in singularities at the corners of the inlet due to the incompatibility between imposed plug flow at the inlet and the no-slip condition for velocity at the walls. At these points, the streamwise flow deceleration is theoretically infinite, and in practice the stress and strain-rate tensors will have large magnitudes at the cells close to these corners, and the magnitude of $\bm{\sigma} \cdot \bm{\sigma} \cdot \dot{\bm{\gamma}} \cdot \dot{\bm{\gamma}} + \dot{\bm{\gamma}} \cdot\dot{\bm{\gamma} } \cdot\ \bm{\sigma} \cdot \bm{\sigma}$ will consequently also be very large. Even with small values of $g_9$, the predicted rate of change of stress in these cells will be large, leading to rapid and unbounded stress growth causing numerical breakdown of the simulation. In practice, it is very difficult or potentially impossible to prevent noise in coefficient channels even when the expected behaviour is that it should identically vanish; such noise has already been observed in irrelevant output channels when the TBNNs with 8 and 9 basis tensors are trained on the Giesekus model (Figures \ref{fig:5} and \ref{fig:6}). It is worthwhile to note that even imposing the fully developed stress and velocity profiles of the Giesekus fluid at the inlet and performing the CFD simulation using the 9 basis TBNN results in instability, due to the incompatibility between the inlet condition and the initial conditions of zero stresses and velocity everywhere.

The TBNN employing eight basis tensors exhibits improved numerical stability by foregoing the basis tensor in $\bm{\sigma} \cdot \bm{\sigma} \cdot \dot{\bm{\gamma}} \cdot \dot{\bm{\gamma}} + \dot{\bm{\gamma}} \cdot\dot{\bm{\gamma} } \cdot\ \bm{\sigma} \cdot \bm{\sigma}$. With this model, we are able to obtain converged simulations describing steady 2D slit flow at $De = 0.5$; however, at $De = 2.5$ the model erroneously predicts unsteady flow and refinement of the mesh and reduction of the time-step leads to numerical divergence; the ground-truth Giesekus model remains steady for these conditions irrespective of mesh refinement. Although the removal of the quartic term obviates the rapid instability onset observed for the nine basis model allowing for stability at low values of $De$, increasing $De$ eventually leads to a numerical instability like in the 9 basis TBNN.

In our tests on the newly proposed TBNN employing four basis tensors, the numerical stability was virtually identical to the ground-truth model, and we are able to obtain converged simulations describing steady flow at least up to $De = 30$. The vastly improved stability of the TBNN employing only 4 basis tensors is an important finding, since it allows quantitative analysis of CFD simulations employing UDE-based constitutive models to be undertaken at non-trivial elastic conditions.

\subsection{Sudden Contraction Flows} \label{subsec:contraction}

The first complex flow to be considered is creeping flow in a 4:1 sudden contraction, which is a classical and widely studied benchmark flow in computational rheology \cite{hassager1988working}. Here, we consider the 2D planar contraction case and also demonstrate the generalisability of the proposed 4 basis model to full 3D flows by considering the flow in a 3D square-square contraction case. One of the key features of these flows is the recirculations which manifest near far corners in the upstream channel at the contraction plane; these are known to be very sensitive to fluid rheology. Their normalised length $\widetilde{X}_R = X_R/h_2$ (where $h_2$ is the downstream channel half-width), has thus been a central focus of past investigations considering the planar case (e.g. \cite{alves_benchmark_2003, afonso_dynamics_2011, choi1988numerical}), and also in the square-square case where it can be measured at the central plane (e.g. \cite{sousa2011effect}).  Here we focus on the flow of fluids obeying a UDE-based constitutive model, trained on Giesekus, ePTT and sPTT fluids under oscillatory shear flow. For both the planar and square-square contraction, the Deborah number is defined based on the downstream conditions, $De = \lambda U_2/h_2$, where $U_2$ is the average velocity in the downstream channel.

\subsubsection{Planar Contraction}

We first consider UDEs trained on Giesekus fluids, which is the same case considered by Lennon et al. \cite{lennon2023scientific}, who reported a UDE simulation in 4:1 planar contraction at $De = 1$ employing the original nine basis TBNN model. In this case, we consider exactly the same numerical and meshing setup as \cite{lennon2023scientific}, with $Re = 0.1$ and $\beta = 1/9$. In our experience employing the nine basis TBNN model in the same case, numerical instability is difficult to avoid even at low $De$. The eight basis TBNN model shows improved stability, although it is still ultimately susceptible to numerical divergence in the low-moderate $De$ regime. In contrast, the stability of the four basis TBNN model proposed here is virtually identical to that of the ground-truth Giesekus model and with this we were able to obtain converged simulations describing steady flow at $De = 20$. This represents a significant improvement upon previous efforts, with the proposed model being capable of unlocking high $De$ flow simulations.  

\begin{figure}[htp]
    \centering
    \includegraphics[width=0.4\textwidth]{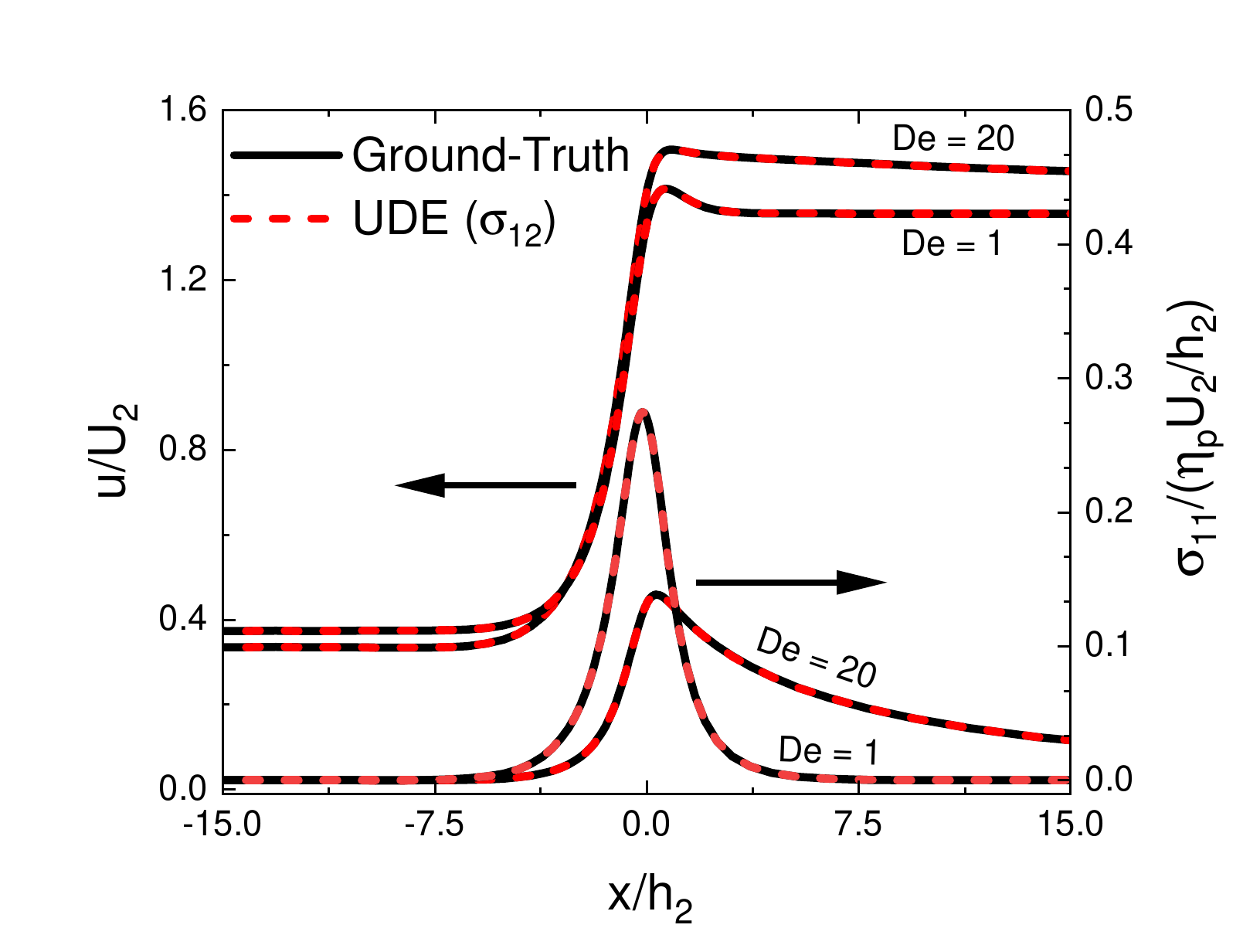}
    \caption{Comparison between the trained UDE and the corresponding ground truth Giesekus fluid $\alpha = 0.4$ streamwise velocity and stress component along the centreline ($y=0$) for the cases $De = 1$ and $De=20$ in a 4:1 planar sudden contraction. Creeping flow is considered with $\beta = 1/9$.}
    \label{fig:17}
\end{figure}

The velocity, stress and pressure fields obtained in the simulation employing the trained UDE are essentially indistinguishable from those with the ground-truth Giesekus model, for all $De$ until numerical stability is lost in both the ground-truth and the UDE cases. Shown in Figure \ref{fig:17}, as an illustrative example, the streamwise velocity and normal stress profiles along the centreline ($y=0$) obtained with the UDE are equally indistinguishable from those of the ground-truth solution at $De = 20$ as they are at $De=1$.

\begin{figure}[htp]
    \centering
    \includegraphics[width=1\textwidth]{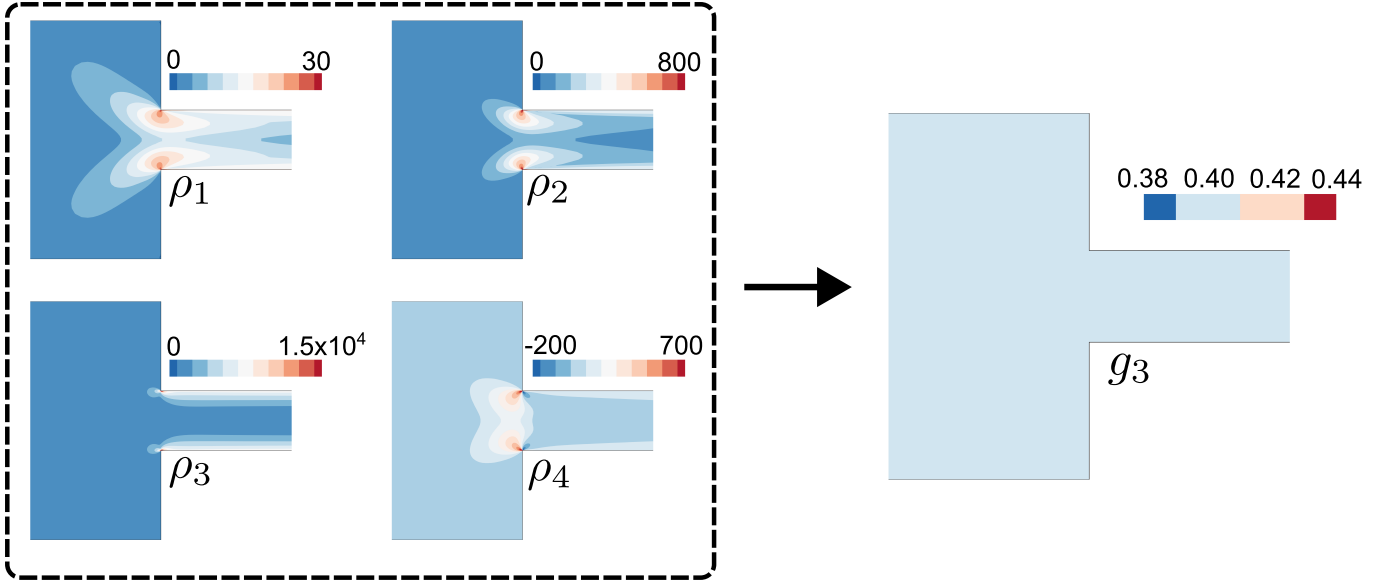}
    \caption{Illustration of the spatial distribution of the inputs, $\rho_i \;(i=1,...,4)$ (left), to the TBNN in the vicinity of the 4:1 contraction plane, and the spatially uniform output $g_3$ (right) at $De =20$ for a UDE trained on a Giesekus fluid $\alpha = 0.4$. Creeping flow is considered with $\beta = 1/9$. }
    \label{fig:18}
\end{figure}

Inspection of the coefficient fields produced by the TBNN for $De = 20$, shows that all of the coefficient fields are spatially uniform, with $g_3 = 0.3995$ (Figure \ref{fig:18}) and all other coefficient fields ($g_1, g_2$ and $g_4$) of the order $10^{-3}$ or less. From this perspective, which illustrates that the UDE universally produces almost the exact Giesekus model throughout the domain (an extension of what was previously observed via direct analysis of the coefficients produced within the training envelope in LAOS only), the almost perfect predictive accuracy of the UDE is easily understood as a natural consequence. It is worthwhile to point out that the inputs to the TBNN (Figure \ref{fig:18}) are much larger than what was encountered during training (in some cases, in the order of thousands of times higher), yet this has virtually no impact on the robustness of the TBNN outputs, which is consistent with expectation given our knowledge of the internal structure of the TBNN for this case.

\begin{figure}[htp]
    \centering
    \includegraphics[width=1\textwidth]{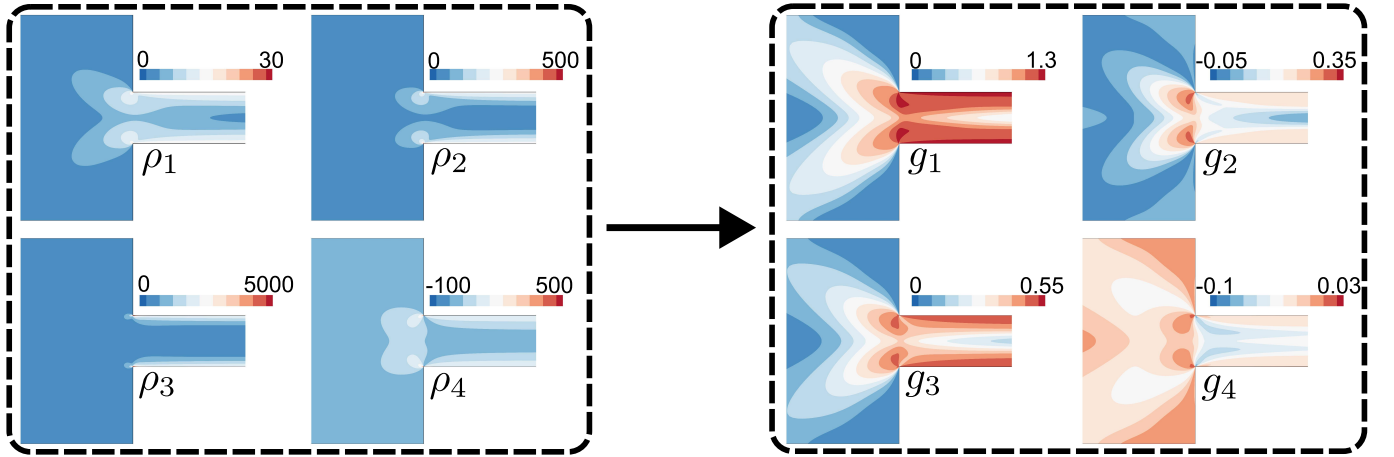}
    \caption{Illustration of the spatial distribution of inputs, $\rho_i \;(i=1,...,4)$ (left), to the TBNN in the vicinity of the 4:1 contraction plane, and the resulting non-uniform TBNN outputs $g_i$ (right) for a steady flow at $De = 10$ for a UDE trained on ePTT fluid with $\epsilon = 0.25, \xi=0$ and exposed to $\sigma_{12}$ only during training. Creeping flow is considered, with $\beta = 1/9$. }
    \label{fig:19}
\end{figure}

\vspace{1em} 
Having established that the UDE trained on the Giesekus model can be deployed on the 4:1 planar sudden contraction with very good predictive accuracy, a natural next step is to assess the performance of a UDE which both has not perfectly captured the underlying rheology, and also learns a complex, variable TBNN representation. To this end, the UDEs trained on the ePTT model ($\epsilon =0.25, \xi=0$) are deployed in creeping flows for varying Deborah numbers in the range $0.1 \leq De \leq 25$. Unlike in the Giesekus case, for simulations with the ePTT-trained UDEs the distribution of TBNN outputs is non-trivial and exhibits spatio-temporal variation over the domain (Figure \ref{fig:19}). In the forthcoming analysis, we consider the UDE trained on the shear component of stress only ($\textnormal{UDE}_{\sigma_{12}}$), and the UDE trained on both the shear component of stress and the first normal stress difference ($\textnormal{UDE}_{\sigma_{12}+N_1}$).

For ground-truth comparison, we refer to the benchmark data of Alves et al. \cite{alves_benchmark_2003}, and we also generate ground-truth data with the existing ePTT model implementation in rheoTool. The agreement between our present ground-truth calculations and the benchmark reference data of \cite{alves_benchmark_2003} is very good for the range of conditions tested, with errors in vortex size below 0.4\% (Figure \ref{fig:21}). 

\begin{figure}[htp]
    \centering
    \includegraphics[width=0.6\textwidth]{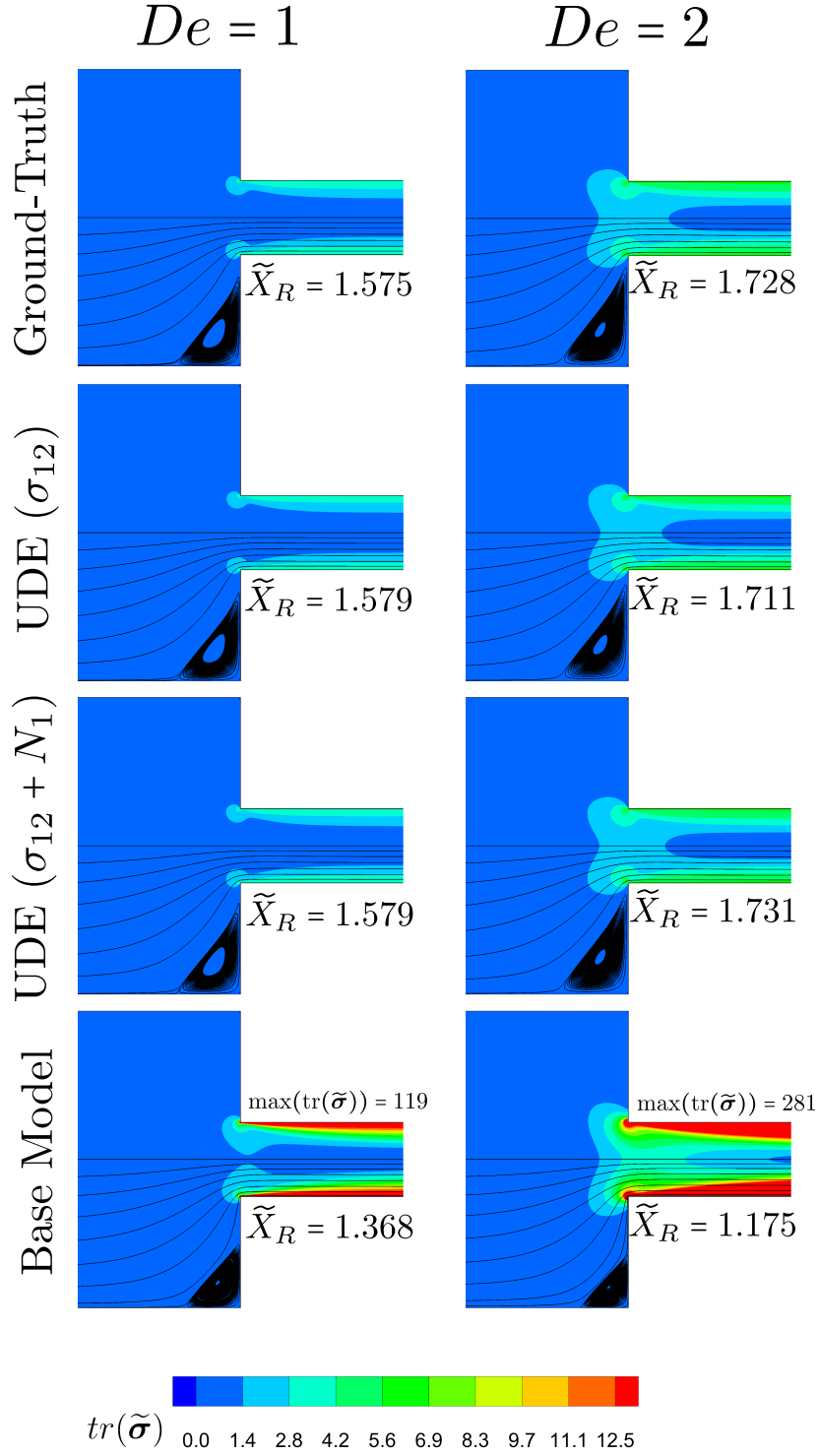}
    \caption{Flow patterns and contour plots of the trace of the stress tensor $\left(\textnormal{tr}(\bm{\widetilde{\sigma}}) = \textnormal{tr}\left( \frac{\bm{\sigma}}{G_0} \right)\right)$ in the 4:1 planar sudden contraction 
    at $De = 1$ and $De = 2$: comparison between the ground-truth constitutive model (ePTT, $\epsilon = 0.25, \xi=0$, $\beta = 1/9 $), the UDE exposed to shear stress only, the UDE exposed to shear stress ($\sigma_{12}$) and first normal stress difference, and the base model (Oldroyd-B model).  }
    \label{fig:20}
\end{figure}

Considering the cases where $De \leq 2$, the entire emergent flow behaviours - including kinematics, stresses and pressure fields -  predicted by the UDEs are essentially indistinguishable from the ground-truth. This is clearly seen in the flow patterns and the contours of the stress tensor shown in Figure \ref{fig:20}, and also in the vortex size shown in Figure \ref{fig:21}. Even though the corner vortex length is sensitive to fluid rheology, the predictions of both UDEs are well within $1\%$ of the ground-truth.

As the Deborah number is increased ($De \geq 5$), the beginnings of small but measurable deviations from the ground-truth can be observed in the stress field and flow kinematics in particular for $\textnormal{UDE}_{\sigma_{12}}$ (Figure \ref{fig:22}). The general pattern, more noticeable for larger $De$ ($De$ = 25), is that the stress magnitudes are over-predicted by the UDEs in comparison to the ground-truth. Nevertheless, the qualitative shapes of the emergent stress contours remain similar to the ground-truth. 

\begin{figure}[htp]
    \centering
    \includegraphics[width=0.35\textwidth]{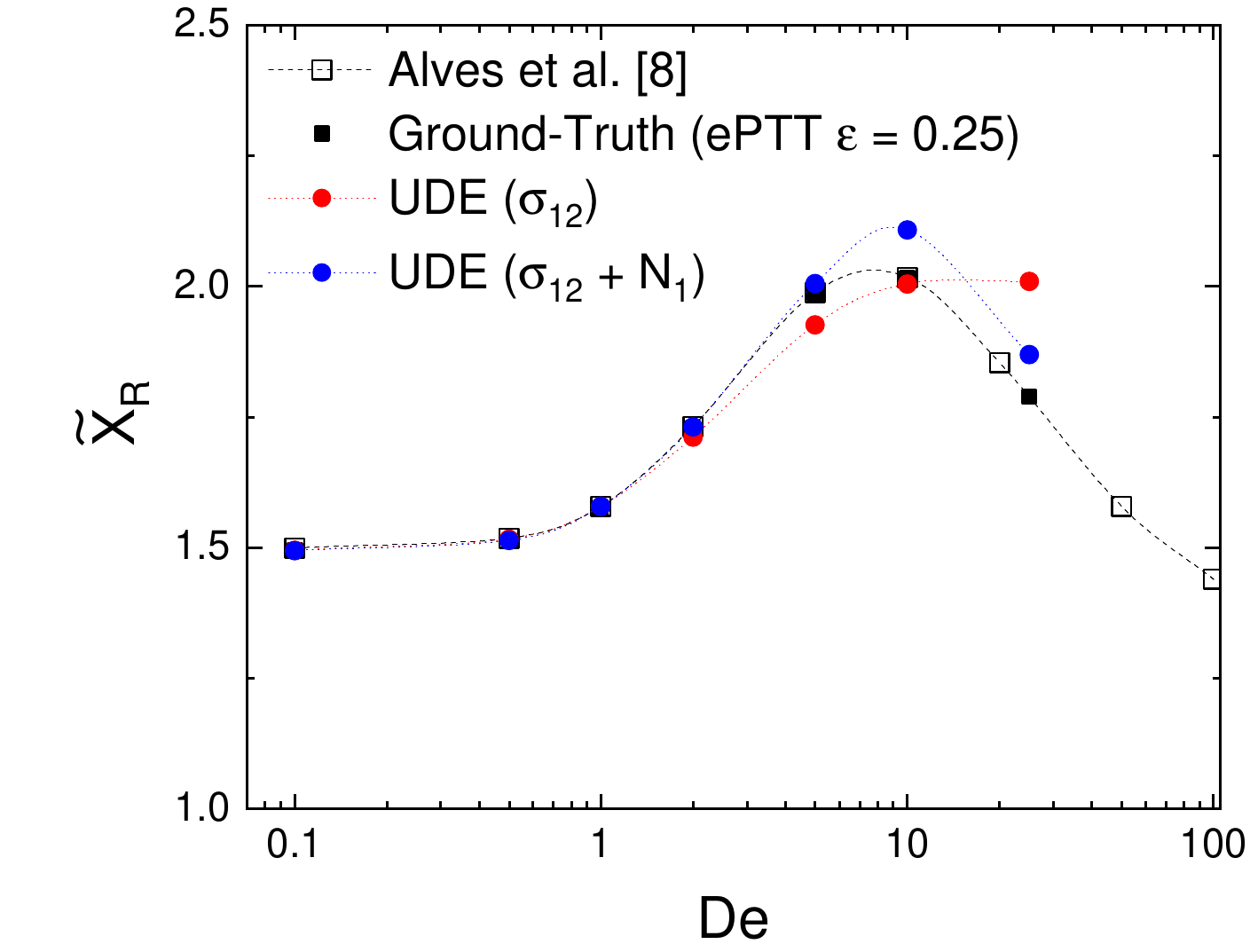}
    \caption{Comparison of the variation of the normalised corner vortex length, $\widetilde{X}_R$, with $De$ in the 4:1 sudden contraction: comparison between the ground-truth constitutive model (ePTT, $\epsilon = 0.25, \xi =0$, $\beta = 1/9 $), the UDE exposed to shear stress only ($\sigma_{12}$), the UDE exposed to shear stress and first normal stress difference ($\sigma_{12}+N_1$) and data from Alves et al. \cite{alves_benchmark_2003}.  }
    \label{fig:21}
\end{figure}

Similarly, the vortex length dependency on $De$ (quantified in Figure \ref{fig:21}) remains similar to the ground-truth up to $De = 10$ for $\textnormal{UDE}_{\sigma_{12}}$ and up to $De = 25$ for $\textnormal{UDE}_{\sigma_{12}+N_1}$, with $\textnormal{UDE}_{\sigma_{12}+N_1}$ capturing a maximum in $\widetilde{X}_R$ as observed in the ground-truth ePTT fluid (which is related to a peak in the Trouton ratio and the recoverable stress \cite{alves_benchmark_2003}). 

The contraction flow inherently contains an extensional component: the flow is purely extensional along the centreline and in much of the domain the mixed kinematics contains a substantial extensional flow component. As $De$ is increased, the extensional response of the UDEs deviates from that of the ground-truth (cf. Figure \ref{fig:15}) and contributes to the differences observed in $\widetilde{X}_R$. Although it is not possible to directly relate the planar extensional viscosity profiles of the UDEs to the Lagrangian unsteady kinematics of the contraction, it is worthwhile to again mention that the failure region for the extensional viscosity of $\textnormal{UDE}_{\sigma_{12}}$ begins at a lower extensional Deborah number than for $\textnormal{UDE}_{\sigma_{12}+N_1}$, and these differences in part account for the qualitative difference in the behaviour of $\widetilde{X}_R$ when $De > 10$.

\begin{figure}[htp]
    \centering
    \includegraphics[width=0.9\textwidth]{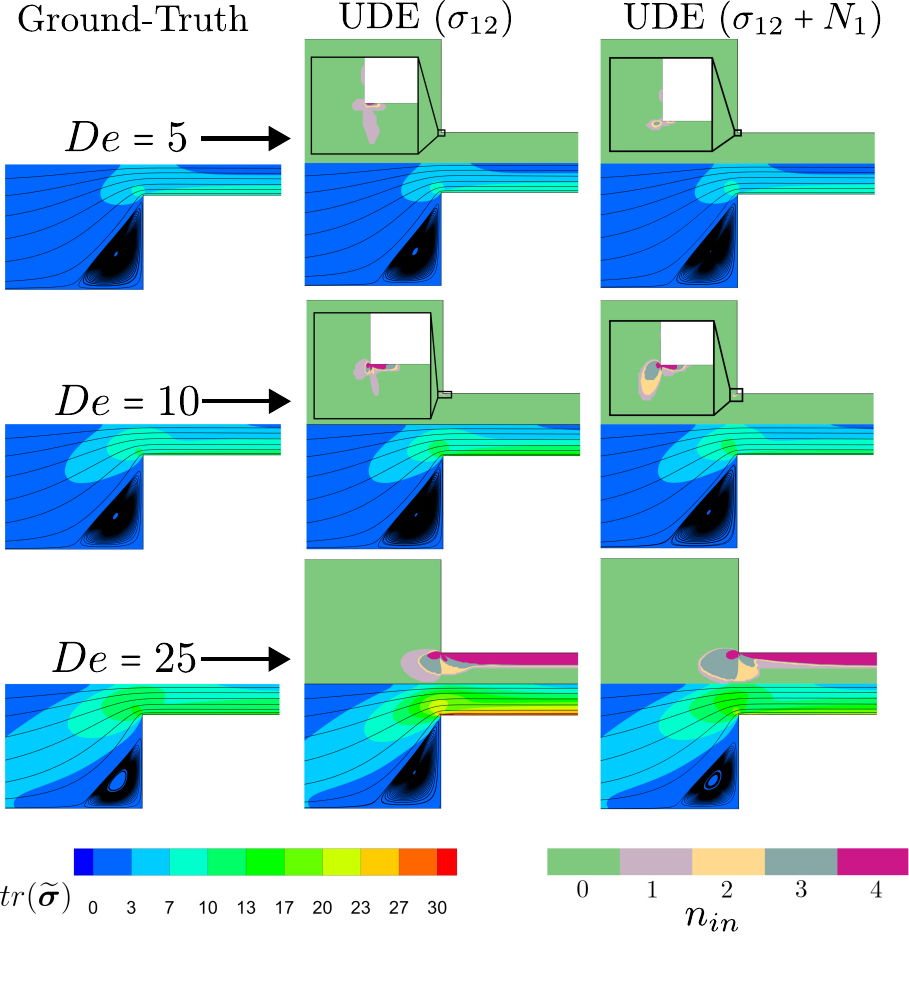}
    \caption{Comparison of the behaviour of the ground-truth (ePTT, $\epsilon=0.25$, $\xi =0$, $\beta = 1/9 $) and UDEs in the 4:1 planar sudden contraction flow at $De = \{5, 10, 25\}$. Streamlines superimposed on the trace of the normalised stress tensor $\textnormal{tr}(\widetilde{\bm{\sigma}})$ (bottom half). Map of the number of TBNN inputs operating in an extrapolative regime, $n_{in}$, for the UDE fluid flows (top half), where $n_{in} = 0$ corresponds to the TBNN operating strictly within the component-wise bounds of the invariants exposed during training, and $n_{in} = 4$ corresponds to every input to the TBNN being outside the bounds of training (extrapolating).}
    \label{fig:22}
\end{figure}

By deploying the UDEs in contraction flows, the models are generalising beyond the LAOS flows on which they were trained. Nevertheless, even if the specific stress-strain-rate states which manifest in the contraction flow were not encountered during training, the TBNN predictions are said to be in the interpolating regime so long as it operates within the component-wise bounds of the invariant space explored during training. It is well established that the extrapolation performance of neural networks is much poorer \cite{haley1992extrapolation}, and it is thus additionally of interest to quantify where and how many numerical inputs lie outside the invariant range spanned by the training data and are thus in an extrapolating regime. Accordingly, for the cases $De=\{5,10,25\}$ we identify and map regions of extrapolation by performing a cell-wise evaluation of the number of TBNN input features, $n_{in}$, operating outside the training domain. The training domain here is defined by the maximum and minimum values of each $\rho_i$ which is observed by the TBNN during training as shown in e.g. Figure \ref{fig:3}. Thus, when $n_{in} = 0$, the TBNN is operating strictly within the component-wise bounds of the invariants exposed during training, and where $n_{in} = 4$ every input to the TBNN is extrapolating. Naturally, the local fluid state near the re-entrant corner reaches a stress-strain-rate state which is beyond the learned invariant space first, as is visible in a small neighbourhood around the re-entrant corner at $De = 5$, and this region increases in size with $De$. At $De = 25$, a significant region around the corner and in the outlet channel are in a completely extrapolating TBNN regime. In the neighbourhood of the re-entrant corner the flow is dominated by solid-body rotation, and in the fully developed region of the outlet channel the flow is pure shear, and so here the deviation from the ground-truth does not arise due to strong deviations in the extensional viscosity, but instead from the combined errors of extrapolation.

Identifying extrapolation regions of the TBNN is critical for ensuring the reliability of simulation results. As $De$ increases, an increasingly large fraction of the domain enters an extrapolating TBNN regime. Consequently, simulations employing the UDEs experience numerical divergence at $De = 100$, despite the ground-truth ePTT model remaining stable even for $De = 10^4$. This spurious divergence arises because a very significant portion of the domain operates in an extrapolating regime; in these regions, the constitutive models predicted by the TBNN are frame-invariant but they do not reflect the true local stress–strain-rate state, generating unphysically large stresses that the numerics cannot balance, analogous to the numerical instabilities of the high Weissenberg number problem. Our result at $De = 25$ demonstrates the robustness of the 4 tensor basis formulation in maintaining numerical stability across regimes of substantial extrapolation in the domain, but reveals an important caveat regarding interpretability because numerical stability is insufficient as a proxy for physical correctness. 

Nevertheless, it is important to point out that the numerical diagnostic proposed here is not ``physics-aware" because it cannot distinguish regions of the flow which, for example, are extensional and thus outside of the training data but remain within the bounds of the invariant space explored during training. Therefore, identifying whether the TBNN operates in a component-wise interpolation regime is necessary but not sufficient for judging the reliability of a numerical simulation, since the TBNN might be extrapolating to new flow dynamics never encountered in training. In practical terms, simulations employing UDEs should therefore monitor this indicator carefully, but cannot rely on it alone to guarantee physical fidelity.

In summary, the deployment of the UDEs trained on an ePTT fluid in the planar sudden contraction has demonstrated encouraging results. At low $De$, the UDE simulations are essentially indistinguishable from the ground-truth, before gradually deviating from the ground-truth whilst retaining numerical stability. This smooth divergence was similarly observed by Shanbhag and Erlebacher \cite{shanbhag2024sparse} in LAOS flows; in the full CFD simulations performed here, the smooth deviation in the emergent flow field can be explained by the deviation of the fluid material properties, particularly the extensional viscosity, from those of the ground-truth. Despite this, the overall trends in flow behaviour with $De$ are remarkably similar to the ground-truth, especially for the UDE also exposed to $N_1$ during training. At sufficiently high $De$, simulations exhibit a HWNP-type numerical instability. Importantly however, the emergence of an extrapolating TBNN regime can occur before numerical instability, and so simulations employing constitutive models described by UDEs must be carefully validated with diagnostics demonstrating that the TBNNs are operating within a suitable regime.

As we have observed, the loss of the fidelity of the extensional viscosity profile in $\textnormal{UDE}_{\sigma_{12}}$ has a significant impact on the predictive accuracy of simulations in the contraction flow. One possible way to therefore improve the accuracy of the predictions is to expose the TBNN to extensional flow during training. One way this could be done is via large amplitude oscillatory extension, recently demonstrated to be experimentally tractable by Recktenwald et al. \cite{recktenwald2025large},  which would directly impart knowledge of the extensional viscosity and potentially correct the deviations which have been observed to manifest at relatively low $\lambda \dot{\epsilon}$.

\subsubsection{3D Square-Square Contraction} \label{subsubsec:3D}

\begin{figure}[htp]
    \centering
    \includegraphics[width=0.7\textwidth]{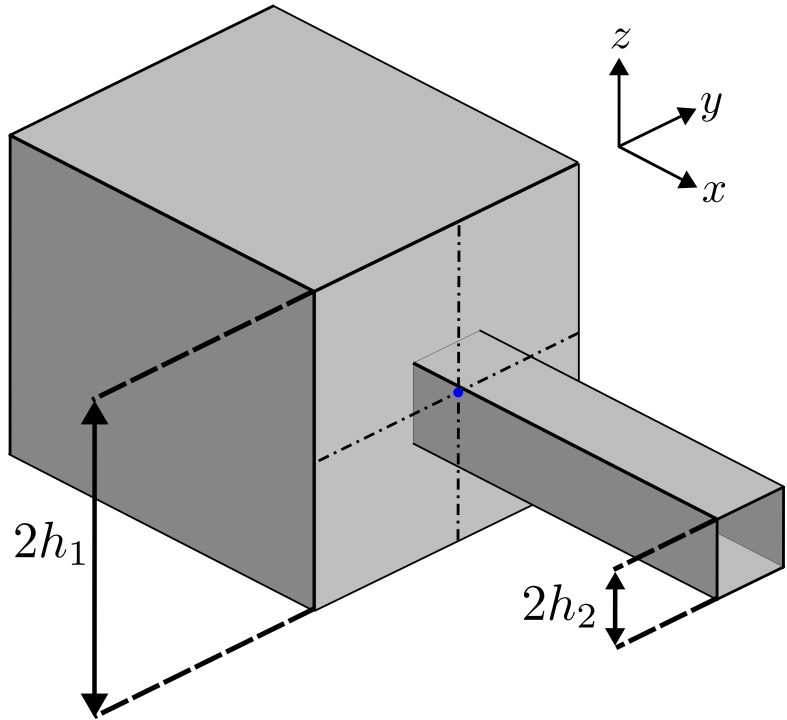}
    \caption{Isometric view of a representative computational domain of the 4:1 square-square sudden contraction in vicinity of the contraction plane. Flow is in the positive $x$-direction. The coordinate system is centred at the contraction plane such that the centre of the contraction plane lies at $x=y=z=0$. }
    \label{fig:23}
\end{figure}

To demonstrate compatibility with 3D flow predictions and test its performance, we deploy a trained UDE in full 3D simulations in a 4:1 square-square contraction (Figure \ref{fig:23}). The contraction ratio is defined as $CR = h_1/h_2 = 4$, where $h_1$ and $h_2$ refer to the half-width of the inlet and outlet channels, respectively. The centre of the contraction plane lies at $x = y = z = 0$

Three simulations in creeping flow conditions are performed at $De = 4$ considering the base model, the ground-truth sPTT model ($\epsilon = 0.25$) and a UDE trained on the sPTT model, exposed to shear stress only. In all cases, a fluid with $\beta = 1/9$ is considered.

\begin{figure}[htp]
    \centering
    \includegraphics[width=0.75\textwidth]{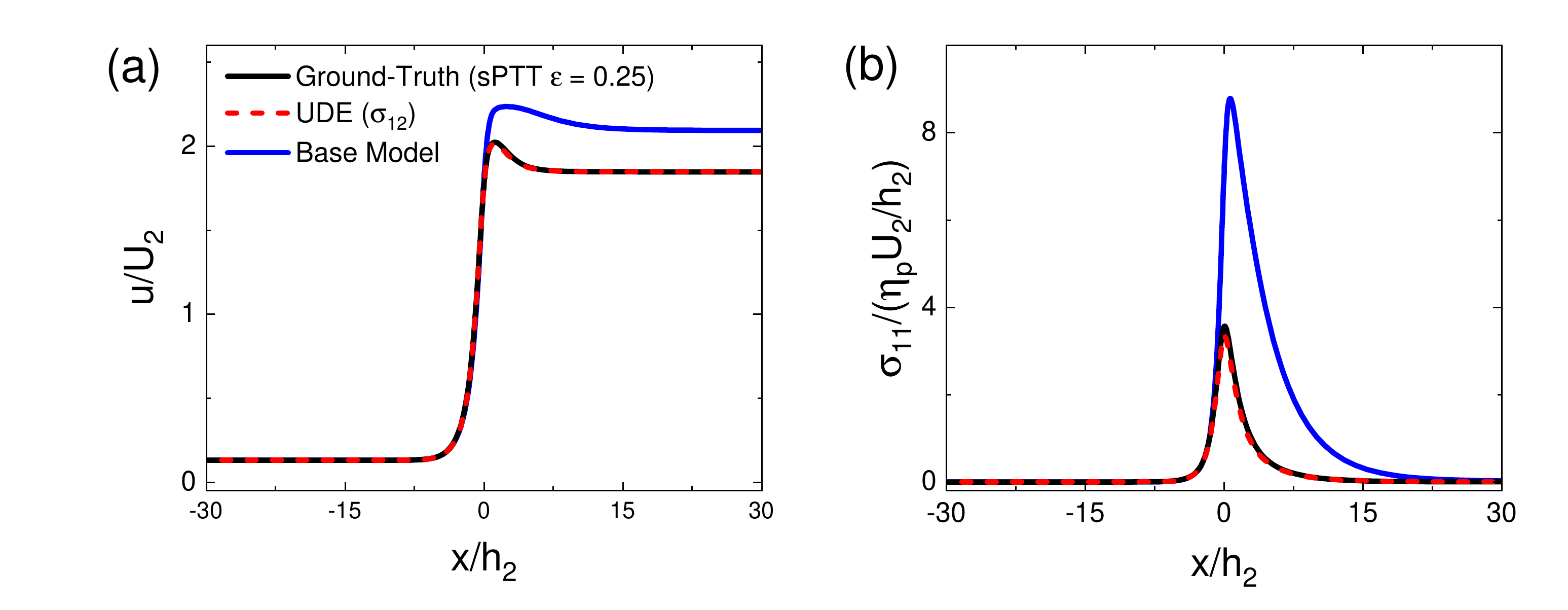}
    \caption{ Comparison of the ground-truth fluid (sPTT $\epsilon=0.25$, $\beta = 1/9$), the UDE trained on shear stress only ($\sigma_{12}$), and the base model (Oldroyd-B) in the 4:1 square-square sudden contraction at $De = 4$ in terms of: (a) the streamwise velocity along the centreline ($y=z=0$); and (b)
    the normal stress profile along the centreline ($y=z=0$).  }
    \label{fig:24}
\end{figure}

\begin{figure}[h!]
    \centering
    \includegraphics[width=0.8\textwidth]{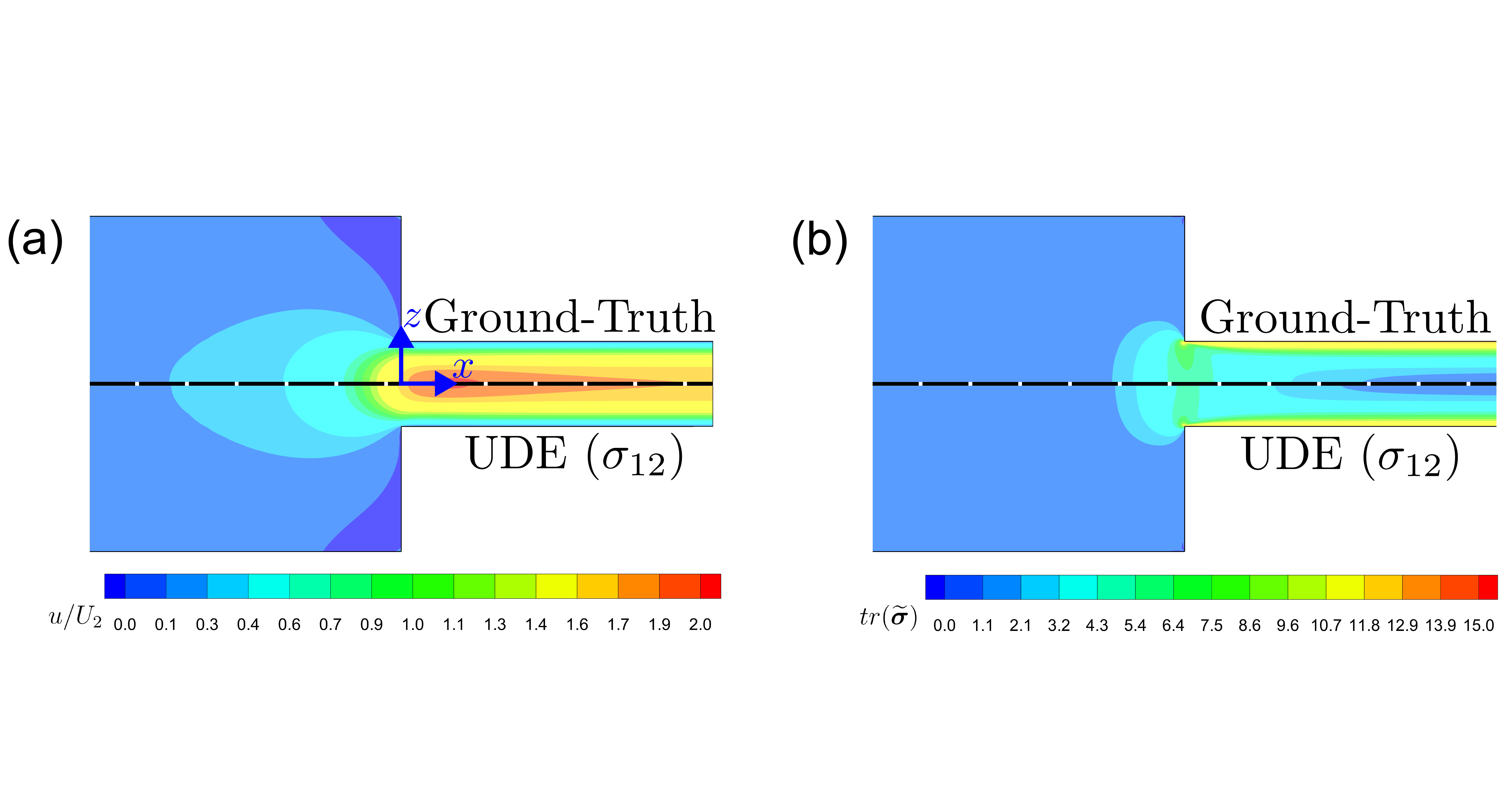}
    \caption{Comparison of the ground-truth (sPTT $\epsilon=0.25$, $\beta = 1/9$) and the UDE trained on shear stress only ($\sigma_{12}$) at $De = 4$ in terms of the: (a) contour plot of the streamwise velocity at a central plane ($y =0$); and (b) contour plot of the trace of the normalised stress tensor at a central plane ($y =0$).}
    \label{fig:25}
\end{figure}

Considering the streamwise velocity and normal stress profiles at the centreline ($y=z=0$), the predictions of the UDE are inline with the ground-truth sPTT model (Figure \ref{fig:24}). Similarly, considering a central plane at $y=0$, the contours of both the streamwise velocity, $u/U_2$, and the trace of the stress tensor, $\textnormal{tr}(\widetilde{\bm{\sigma}})$, are closely approximated by the UDE (Figure \ref{fig:25}). This illustrates that the correction term, though purely motivated via 2D considerations and trained purely on 2D flows, retains the potential to accurately predict 3D flows in non-trivial elastic conditions.

\subsection{Planar Cross-Slot Flow} \label{subsec:cross-slot}

In this section, we assess the performance of the TBNNs in a computational benchmark flow that exhibits a stagnation point and a strong extensional component \cite{haward2012optimized}. For this purpose, we simulate the flow of sPTT-trained TBNNs through a planar cross-slot, which is expected to exhibit a purely elastic instability resulting in a supercritical bifurcation to steady asymmetric flow above a critical value of the Deborah number \cite{poole2007purely}.
Deployment of trained UDEs in the planar cross-slot configuration can therefore expose its behaviour in regions of instability, as well as more generally in regions where accurate stress prediction is critical. 

We consider a cross-slot with inlet and outlet arms of width $w$ and  length $25w$, which is found to be sufficient for full development of the velocity and stress profiles for the conditions examined here. As the goal of the present work is to compare the behaviour of trained UDE's and the corresponding ground-truth constitutive model rather than to provide or validate benchmark data, we opt to employ a mesh with an intermediate level of refinement corresponding to Mesh M2 of Cruz et al. \cite{cruz2014new} ($\Delta x_{min}/w = \Delta y_{min}/w = 0.01$). In this geometric configuration, the Deborah number is defined as $De = \lambda U/w$ where $U$ is the average flow velocity in each arm.

For the training we use the sPTT fluid with $\epsilon = 0.02$ and, as in the previous section, we consider UDEs trained on either the shear component of stress alone ($\textnormal{UDE}_{\sigma_{12}}$), or trained on both the shear component and the first normal stress difference ($\textnormal{UDE}_{\sigma_{12}+N_{1}}$). For all CFD simulations, we consider creeping flow for a fluid with $\beta = 1/9$.

\begin{figure}[htp]
    \centering
    \includegraphics[width=.9\textwidth]{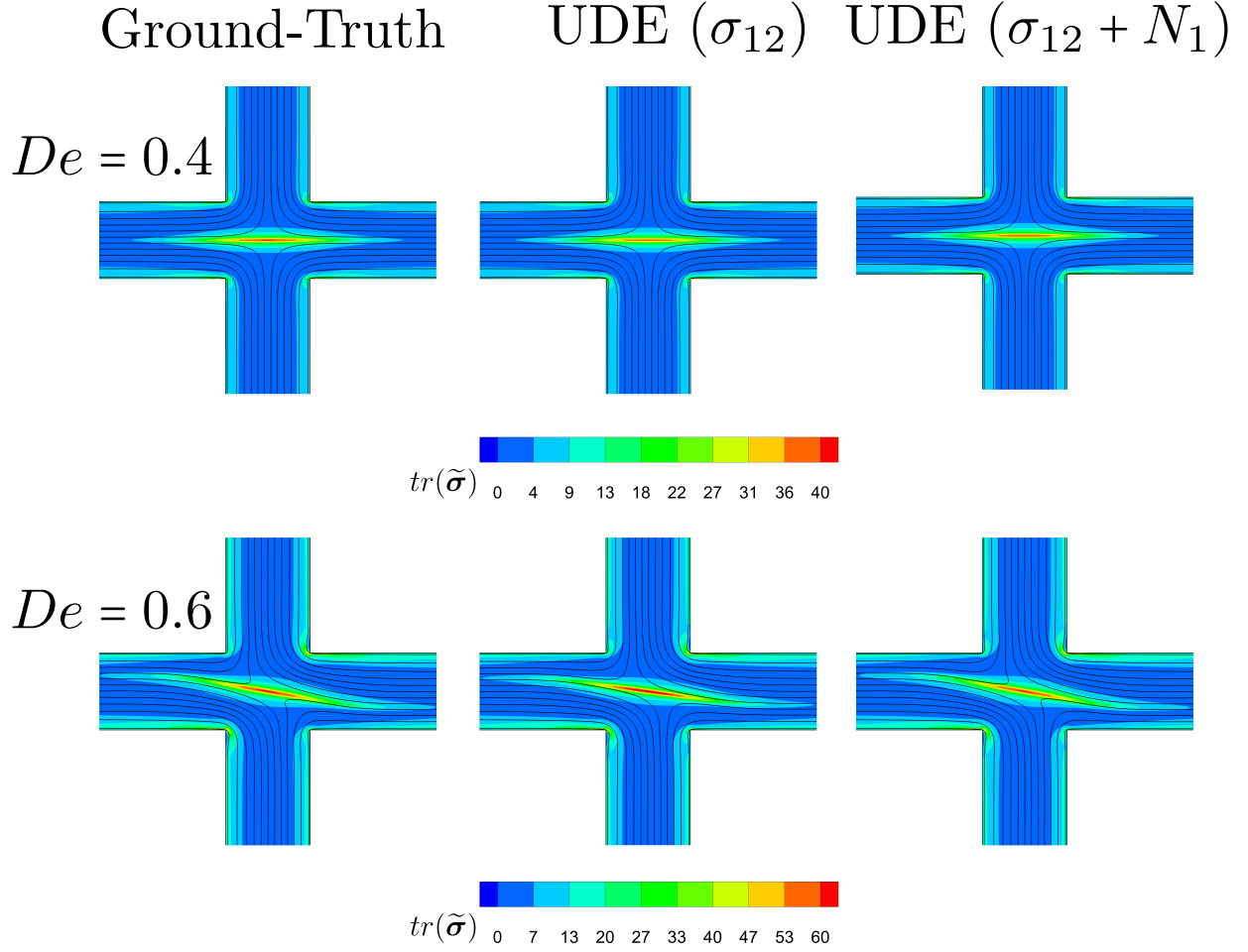}
    \caption{Streamlines and contour plots of the trace of the normalised viscoelastic extra-stress tensor in a planar cross-slot geometry, $\textnormal{tr}(\widetilde{\bm{\sigma}})$, for fully symmetric flow ($De = 0.4$), and steady asymmetric flow  ($De = 0.6$) obtained below and above the critical conditions for the onset the elastic instability, respectively. Comparison between the ground-truth constitutive model (sPTT, $\epsilon = 0.02$, $\beta = 1/9$), the UDE exposed to shear stress only ($\sigma_{12}$), and the UDE exposed to shear stress and first normal stress difference ($\sigma_{12} + N_1$), under creeping flow conditions.}
    \label{fig:26}
\end{figure}

Qualitatively, both trained UDEs are able to capture the steady asymmetry as shown in  Figure \ref{fig:26}. We can observe that the emergent flow patterns and stress fields of the UDEs at $De = 0.4$ and $De =0.6$, respectively corresponding to conditions before and after the onset of the elastic instability, are in very reasonable agreement with the ground-truth. At $De = 0.4$, the fields of $\textnormal{tr}(\widetilde{\bm{\sigma}})$ obtained using the UDEs and the groud truth constitutive modelare almost indistinguishable from one another. At $De = 0.6$, well beyond the conditions where the onset of steady asymmetric flow is expected in the ground-truth sPTT model (expected at a critical Deborah number $De_c = 0.508$ according to the benchmark data of \cite{cruz2014new}), we note some differences in the stress fields. Although both UDEs predict a steady asymmetry, the streamlines of the $\textnormal{UDE}_{\sigma_{12}}$
indicate a lower degree of asymmetry, and the stress state near the geometric centre is both quantitatively larger ($\textnormal{tr}(\widetilde{\bm{\sigma}})$ at the geometric centre is 10\% larger than the ground-truth for $\textnormal{UDE}_{\sigma_{12}}$, while for $\textnormal{UDE}_{\sigma_{12+N1}}$ it is 1\% smaller than the ground-truth) and spatially encompasses a larger region of the flow. The stress states at the sharp corners are also both larger in size and magnitude. In comparison, the streamlines corresponding to the UDE exposed to both shear stress and $N_1$ indicate a stronger degree of asymmetry in line with the ground-truth.

\begin{figure}[htp]
    \centering
    \includegraphics[width=0.5\textwidth]{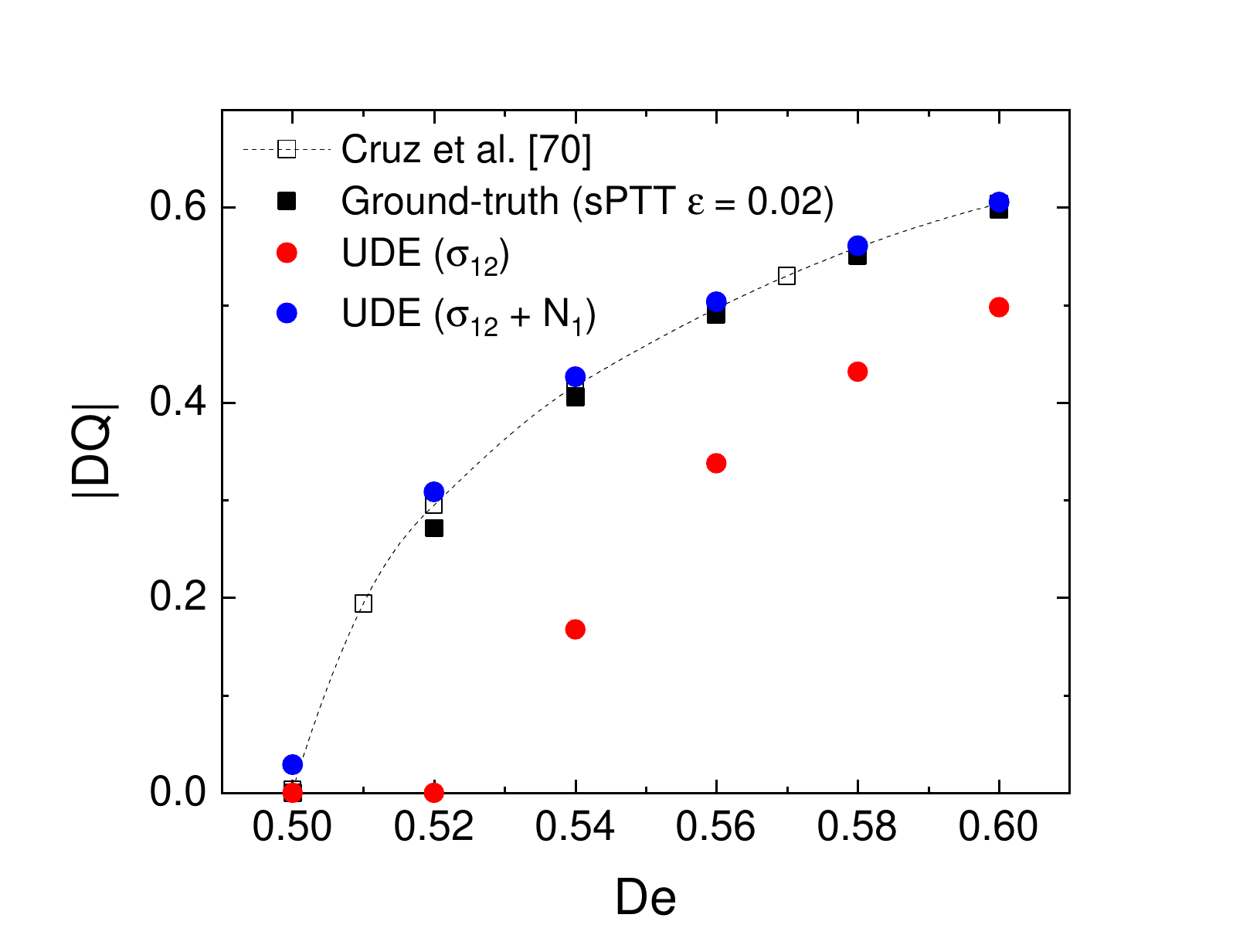}
    \caption{Variation in the magnitude of the flow asymmetry parameter, $|DQ|$, with $De$ for the creeping flow in a planar cross-slot. Comparison between the ground-truth constitutive model (sPTT, $\epsilon = 0.02$, $\beta = 1/9$), the UDE exposed to shear stress only ($\sigma_{12}$), the UDE exposed to shear stress and first normal stress difference ($\sigma_{12} + N_1$), and the benchmark data of Cruz et al. \cite{cruz2014new}. }
    \label{fig:27}
\end{figure}

For a more quantitative analysis of the onset and growth of the asymmetry, the flow asymmetry parameter $DQ$ is plotted in Figure \ref{fig:27} as a function of $De$ in the neighbourhood of the instability. $DQ$ is defined as 

\begin{equation} \label{eq:20}
   DQ = \frac{q_2 - q_1}{q_2 + q_1}
\end{equation}

\noindent considering that the total flow rate per unit depth in each inlet channel, $q_1 + q_2$, divides into $q_1$ going through one of the outlets and $q_2$ going through the other outlet. In the perfectly symmetric case, the flow from one inlet channel divides evenly into both outlet channels and $DQ= 0$; in the limiting case of all the flow from one inlet being diverted through one outlet channel then $|DQ| = 1$.

We compare the two TBNNs results with both the results of our simulations with the real SPTT constitutive model and the results reported by Cruz et al. \cite{cruz2014new} in an equivalent mesh for validation. The supercritical bifurcation (in terms of onset and growth of the bifurcation) captured in our simulations with both the SPTT and with the UDE exposed to both $\sigma_{12}$ and $N_1$ is in very good agreement with the literature \cite{cruz2014new}. Quantitatively, the $\textnormal{UDE}_{\sigma_{12}+N_{1}}$ predicts a slightly earlier transition to asymmetry than the ground-truth; at $De=0.50$, the ground truth flow is symmetric while for the $\textnormal{UDE}_{\sigma_{12}+N_{1}}$ the asymmetry parameter is small but already non-zero ($|DQ| = 0.029$). In contrast, the UDE trained on the shear component alone ($\textnormal{UDE}_{\sigma_{12}}$), exhibits a delayed bifurcation, predicting a steady-symmetric flow at $De = 0.52$, and asymmetric flow at $De = 0.54$. 
To explain this behaviour from a physical point of view, we again consider the emergent material properties of the UDEs. In particular, the flow curve of the $\textnormal{UDE}_{\sigma_{12}}$ resembles that of an sPTT fluid with a larger solvent viscosity ratio (cf. Figure \ref{fig:14}). Increases in solvent viscosity ratio have been shown to shift the onset of asymmetric flow to higher values of $De_c$ across a range of geometrical configurations \cite{rocha2009extensibility} (for sufficiently high viscosity ratios, the steady asymmetric regime may even be suppressed, leading instead to a direct transition to time-dependent flow).

 Overall, the cross-slot represents a more challenging flow scenario for the UDEs to  reproduce accurately than the contraction flow, thanks to the sensitivity of the instability to small changes in the fluid state. Nevertheless, despite the noted quantitative differences between the UDEs predictions and the ground-truth, there is still remarkable qualitative similarity in the general structure of the bifurcation.

\section{Conclusion and Future Work}

In this work, we have integrated the UDE and TBNN constitutive modelling approach into a generalised open-source finite volume solver. Taking into consideration the two-dimensional nature of rheometric flow data, we have derived a reduced set of basis tensors (moving from the 8 or 9 used in previous works \cite{lennon2023scientific, rodrigues2025finding} to 4 in the current work). This change imparts vastly improved stability during training and, critically, in the context of numerical simulations, allowing the simulation of flows at high Deborah numbers (e.g. in  case of the planar contraction flow, with the reduced basis we can achieve $De$ nearly two orders of magnitude higher than with the 8 or 9 basis tensor formulations).

Having trained the UDEs on a variety of non-linear viscoelastic constitutive models, we characterised the TBNN behaviour within the range of input features encountered during training and directly observe the learned TBNN representation. We find that the TBNNs trained on the Giesekus and Johnson-Segalman models recover the true model governing the rheology of the training data, since the target function to be learned by the TBNN coincides with a constant function in one of the integrity basis tensors. In all of the other instances, the target function involves modulation of the local stress state to reproduce the ground-truth constitutive model response, and in these cases the learned tensor basis representation captures an alternative behaviour to the true underlying rheology.

We presented the key material properties of the fluids described by the trained UDEs, namely the shear viscosity ($\eta$) and the first normal stress difference coefficient ($\psi_1$), as well as the extensional viscosity ($\eta_E$). In general, the shear-dependent properties are well reproduced for Deborah numbers below the maximum considered during training, and deviating slightly at Deborah numbers beyond this maximum. The accuracy of the extensional viscosity profile depends more significantly on the training fluid, although the general trend is that the profile is accurately reproduced in the low extensional Deborah regime before significantly deviating from the ground-truth. The introduction of $N_1$ during training has little effect on the recovered shear viscosity profile, but produces a considerable  improvement in the extensional viscosity profile.

One of the key goals of this work was to ascertain the performance of trained UDEs in complex flows using CFD-based analysis, especially where the UDE representation is also complex. To this end, in addition to the Giesekus-trained UDE, we deployed UDEs trained on ePTT and sPTT fluids in three benchmark flows, namely the 4:1 planar/square-square sudden contraction and the planar cross-slot. In general, the flows predicted by the UDEs qualitatively resemble the ground-truth with very reasonable accuracy, both in terms of the emergent stress fields and kinematics. 
The UDEs solely exposed to shear stresses in LAOS measurements are capable of reproducing the trends of benchmark variables, namely the corner vortex length in contraction flows, and the flow asymmetry parameter in cross-slot flows. The emergent behaviour observed in the complex UDEs deployed in the present work is in very good agreement, to the point of indistinguishability, at low $De$ in both 2D and fully 3D flows.
This has demonstrated that using a reduced-dimensional tensor basis formulation and characterising a constitutive model based on LAOS measurements can be sufficient for deployment in complex flow problems in both 2D and 3D; this remains true even in cases where the UDE has not recovered the parsimonious ground-truth model. By exposing the UDE to $N_1$ during training, we observe enhanced quantitative accuracy to higher values of $De$, and in the contraction it also positively influenced the qualitative behaviour of the corner vortex in the higher $De$ regime, bringing it inline with the behaviour of the ground-truth fluid. In addition, we introduced a diagnostic variable to identify regions of the computational domain where the TBNN is operating in an extrapolating regime, and demonstrated that the region of extrapolation grows in size with increasing $De$, leading to reduced physical accuracy and eventually spurious numerical instability associated with decoupling of the TBNN correction from the local stress-strain-rate state. 

In the present work, we have clearly demonstrated the potential of TBNNs in their most restrictive case, where they are exposed solely to shear stress data in LAOS. Even in this scenario, the TBNN correction approach can produce a UDE such that complex flow simulations can be undertaken with a high degree of accuracy within a range of Deborah numbers. As we have already seen, supplying more information leads to a UDE response which more precisely represents the ground-truth fluid, even when that information is also obtained from pure shear flow. Although not explored here, we anticipate that the incorporation of novel flow protocols, especially those which encode information about extensional flow responses, will lead to further improved UDE responses. It is not unrealistic to suggest that with this, it may be possible to obtain a trained UDE which is able to accurately capture complex flow responses at high $De$.

\section*{Acknowledgements}

The authors wish to acknowledge EPSRC PhD Studentship under DTP 2224 EP/W524670/1, national funds through FCT/MECI: CEFT, UID/00532/2026 (\url{https://doi.org/10.54499/UID/00532/2025}) and UID/PRR/00532/2025 (\url{https://doi.org/10.54499/UID/PRR/00532/2025}) and ALiCE, LA/P/0045/2020 (\url{https://doi.org/10.54499/LA/P/0045/2020}). This work was supported by computational resources provided by MareNostrum and the Barcelona Supercomputing Center under project epor17 (\url{https://www.bsc.es/marenostrum}), and the Isambard-AI National AI Research Resource (AIRR) through project AIRR-GW-2026-01-01. Isambard-AI is operated by the University of Bristol and is funded by the UK Government’s Department for Science, Innovation and Technology (DSIT) via UK Research and Innovation; and the Science and Technology Facilities Council [ST/AIRR/I-A-I/1023]. We would like to thank M Fossati for server time and G McKinley and K Lennon for helpful discussions. 

\section*{Data Availability}

The data that support the findings of this study are available upon reasonable request.

\printbibliography[title=References]

\end{document}